%% file: ANA-EXOT-2019-19-PAPER.tex
\documentclass[cernpreprint,atlasdraft=true,texlive=2016,UKenglish,texmf]{atlasdoc}
 
\usepackage{atlaspackage}
\usepackage{atlasbiblatex}
 
\usepackage{atlasphysics}
 
\graphicspath{{logos/}{figures/}}
 
\usepackage{ANA-EXOT-2019-19-PAPER-defs}
\usepackage{graphicx}
\usepackage{epstopdf}
\usepackage[bottom]{footmisc}
\AtlasTitle{Search for pair production of scalar leptoquarks decaying into first- or second-generation leptons and top quarks in proton--proton collisions at $\sqrt{s}$ = 13~TeV with the ATLAS detector
}
 
\AtlasAbstract{
A search for pair production of scalar leptoquarks, each decaying into either an electron or a muon and a top quark, is presented.
This is the first leptoquark search using ATLAS data to investigate top-philic cross-generational couplings that could provide explanations for recently observed anomalies in $B$ meson decays.
This analysis targets high leptoquark masses which cause the decay products of each resultant top quark to be contained within a single high-$p_{\mathrm{T}}$ large-radius jet.
The full Run 2 dataset is exploited, consisting of 139~fb$^{-1}$ of data collected from proton--proton collisions at $\sqrt{s}=13$~$\mathrm{TeV}$ from 2015 to 2018 with the ATLAS detector at the CERN Large Hadron Collider.
In the absence of any significant deviation from the background expectation, lower limits on the leptoquark masses are set at 1480~$\mathrm{GeV}$ and 1470~$\mathrm{GeV}$ for the electron and muon channel, respectively.
}
 
\author{The ATLAS Collaboration}

\AtlasJournal{EPJC}

\PreprintIdNumber{CERN-EP-2020-160}
 
\arXivId{2010.02098}
 
\HepDataRecord{96389}
 
\AtlasJournal{Eur.\ Phys.\ J.\ C.}
\AtlasJournalRef{\EPJC 81 (2021) 313}
\AtlasDOI{10.1140/epjc/s10052-021-09009-8}

\hypersetup{pdftitle={ATLAS document},pdfauthor={The ATLAS Collaboration}}
 
\begin{document}
 
\maketitle

\section{Introduction}
\label{sec:intro}

The quark and lepton sectors of the Standard Model (SM) are interestingly similar, motivating one to hypothesize a fundamental symmetry between the two sectors. Such a symmetry can be found in many grand unified theories, such as grand unified SU(5)~\cite{su5}, the Pati--Salam model based on SU(4)~\cite{patisalam}, or R-parity-violating (RPV) supersymmetry (SUSY) models~\cite{rpv_susy}. These models predict a new class of bosons carrying both lepton and baryon number, called leptoquarks (LQs). LQs are hypothetical colour-triplet bosons which couple directly to quarks and leptons. They can be of either scalar or vector nature, and carry fractional electric charge. The production cross section of vector LQs could be enhanced relative to that of scalar LQs due to the existence of a massive gluon partner in the minimal set of vector companions~\cite{vectorLQ_minimal}.
 
LQs have recently gained attention as they provide an attractive explanation of the recent hints of possible lepton-flavour-universality violation from the observed $B$ meson decay anomalies in BaBar~\cite{babar}, Belle~\cite{belle} and LHCb~\cite{lhcb1,lhcb2,lhcb3}.
Single scalar ($\text{S}_{3}$) or vector ($\text{U}_{3}$) LQ triplet models~\cite{lq_solution_tltl1}, as well as a mixed model of a doublet and a singlet LQ ($\text{R}_{2}+\tilde{\text{U}}_{1}$)~\cite{lq_solution_tltl2} are possible solutions to the flavour-changing neutral current $B$ anomaly.
LQs are also motivated by a long-standing deviation from the SM in the anomalous muon magnetic dipole moment measured with the E821 experiment at Brookhaven National Laboratory~\cite{gminus2,e821}.
At the Large Hadron Collider (LHC)~\cite{lhc_machine}, LQs could be produced in pairs, or singly in association with a lepton.
 
\begin{figure}[!hpb]
\begin{center}
\includegraphics[width=0.84\columnwidth]{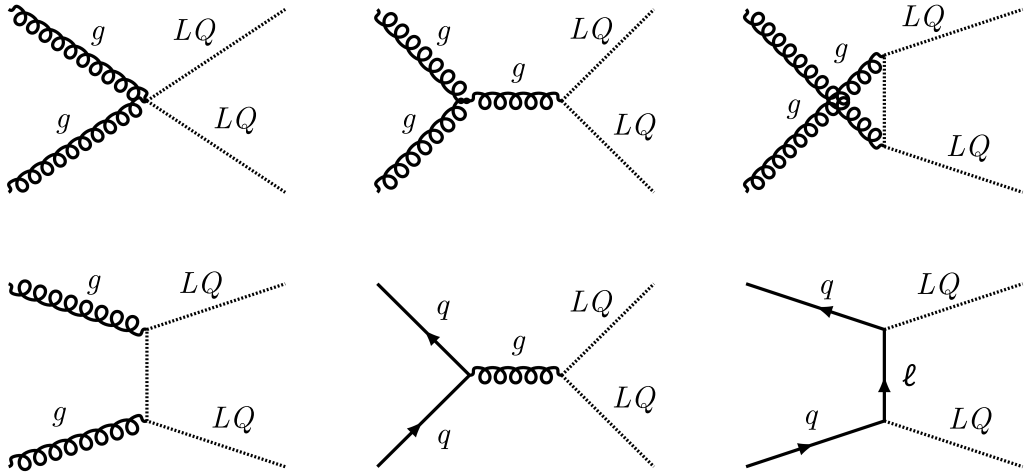}
\caption{The lowest-order Feynman diagrams for LQ pair production. In this paper, the t-channel lepton exchange diagram is ignored.}
\label{fig:LQ_feynman}
\end{center}
\end{figure}

This analysis targets LQ pair production, which is dominated by strong interactions and largely insensitive to the Yukawa coupling at a LQ--lepton--quark vertex. The lowest-order Feynman diagrams are shown in Fig.~\ref{fig:LQ_feynman}. Gluon-initiated processes dominate for LQ masses less than 1.5~\TeV. The t-channel lepton exchange process contributes to the cross section at the 10\% level, and is thus neglected in this analysis~\cite{scalarLQ_LHC,scalarLQmodel1,scalarLQmodel2,scalarLQmodel3}. Only scalar LQ production is considered because this is less model dependent than vector LQ production. The LQ--lepton--quark couplings are determined by two parameters: a model parameter $\beta$, that controls the branching ratio into charged leptons or neutrinos, and the coupling parameter $\lambda$. The coupling to charged leptons is given by $\sqrt{\beta}\lambda$, and the coupling to neutrinos by $\sqrt{1-\beta}\lambda$.
 
Most previous searches have assumed that leptoquarks couple to quarks and leptons of the same generation. Recently, there have been dedicated searches at the LHC for LQ pair production in the $\text{LQ}\rightarrow q\ell$, $\text{LQ}\rightarrow c\ell$, $\text{LQ}\rightarrow b\ell$, and $\text{LQ}\rightarrow t\tau$ channels using the full Run 2 proton--proton ($pp$) collision dataset collected at $\sqrt{s}=13$~\TeV{}~\cite{EXOT-2019-13,Aad:2021rrh}.
The results presented here pertain to the search for cross-generational leptoquarks with decays into a top quark and an electron or a top quark and a muon, in which both top quarks decay hadronically.
It is optimized for LQ masses larger than 1~\TeV, for which the top quarks tend to be boosted.
Therefore, the signature considered is a pair of same-flavour opposite-sign leptons and a pair of large-radius (large-$R$) jets.
Simultaneous couplings of LQs to the first- and second-generation leptons are tightly constrained by the measurements of rare lepton-flavour-violating decays~\cite{LQ_FC_Constraints}, and thus not considered in this paper.
A boosted decision tree (BDT) approach, based on kinematic variables and jet substructure variables, is applied to classify events as originating from the signal or background processes in the signal region.
Dedicated control regions are constructed to control the normalization of the dominant backgrounds: \ttbar{} and \zjets{} production.
The extraction of the signal strength is performed through a simultaneous likelihood fit to the BDT discriminant distribution and the control region yields.
The LQ$\rightarrow t\mu$ and LQ$\rightarrow te$ channels have not been examined previously in ATLAS. The CMS Collaboration has published a search using 35.9~$\text{fb}^{-1}$ of data collected in 2015-2016 that excluded masses below 1420~\GeV\ for scalar LQs decaying exclusively into $t\mu$~\cite{cms_lq}.
 
\section{ATLAS detector}
\label{sec:detector}
 
The ATLAS detector~\cite{PERF-2007-01,ATLAS-TDR-2010-19,PIX-2018-001} at the LHC is a multipurpose particle detector with a
forward--backward symmetric cylindrical geometry that covers nearly the entire solid angle
around the collision point. It consists of an inner detector (ID)
surrounded by a thin superconducting solenoid providing a 2~T axial magnetic
field, electromagnetic and hadronic calorimeters, and a muon spectrometer.
The inner detector covers the pseudorapidity range\footnote{
The ATLAS Collaboration uses a right-handed coordinate system with its origin at the nominal
interaction point (IP) in the centre of the detector
and the $z$-axis along the beam pipe. The $x$-axis points from the IP to the
centre of the LHC ring, and the $y$-axis points upwards. Cylindrical coordinates
$(r,\phi)$ are used in the transverse plane, $\phi$ being the azimuthal angle
around the beam pipe. The pseudorapidity is defined in terms of the polar angle
$\theta$ as $\eta=-\ln\tan(\theta/2)$. Angular distance
is  measured in units of $\Delta R \equiv  \sqrt{(\Delta\eta)^2 + (\Delta\phi)^2} $.}
$|\eta| < 2.5$. It consists of a silicon pixel detector, including the insertable
B-layer installed after Run 1 of the LHC, and a silicon microstrip detector surrounding the pixel detector,
followed by a transition radiation straw-tube tracker.
Lead/liquid-argon sampling calorimeters provide electromagnetic
energy measurements with high granularity and a steel/scintillator-tile
hadron calorimeter covers the central pseudorapidity range ($|\eta| < 1.7$).
The endcap and forward regions are instrumented with liquid-argon calorimeters for
both the electromagnetic and hadronic energy measurements up to $|\eta| = 4.9$.
The outer part of the detector consists of a muon spectrometer (MS) with high-precision tracking chambers for coverage up to $|\eta| = 2.7$, fast detectors for triggering over $|\eta| < 2.4$, and three large superconducting toroid magnets with eight coils each.
The ATLAS detector has a two-level trigger system to select events for offline analysis~\cite{TRIG-2016-01}.

\section{Data and simulation samples}
\label{sec:samples}
 
The data utilized in this search correspond to $139$~\ifb{} of integrated luminosity from \pp collisions at $\sqrt{s}=13$~\TeV{} collected with the ATLAS detector. Only data collected during stable beam conditions with all ATLAS detector subsystems operational are considered.
 
Simulated events with pair-produced scalar LQs were generated at next-to-leading order (NLO) in quantum chromodynamics (QCD) with \MGMCatNLOV{2.6.0}~\cite{Madgraph} using the LQ model of Ref.~\cite{scalarLQmodel1} that adds parton showers to previous fixed-order NLO QCD calculations~\cite{scalarLQmodel2,scalarLQmodel3}, and the NNPDF3.0nlo~\cite{nnpdf30} parton distribution function (PDF) set with \alphas$=0.118$. \MADGRAPH was interfaced with \PYTHIAV{8.230}~\cite{Pythia82} using the A14 set of tuned parameters (tune)~\cite{A14} and the NNPDF2.3lo set of PDFs~\cite{Ball:2012cx} for the underlying-event description, parton showering, and hadronization. Matching of the matrix element with parton showering was performed following the CKKW-L prescription~\cite{ckkw}, with a matching scale set to one quarter of the leptoquark mass. The LQ pair-production cross sections were obtained from the calculation of direct top-squark pair production, as they are both massive, coloured, scalar particles with the same production modes, computed at approximate next-to-next-to-leading order (NNLO) in QCD with resummation of next-to-next-to-leading logarithmic (NNLL) soft gluon terms~\cite{stop-xs1,stop-xs2,stop-xs3,stop-xs4}. The cross sections do not include lepton t-channel contributions, which are neglected in Ref.~\cite{scalarLQmodel1} and may lead to corrections at the 10\% level~\cite{scalarLQ_LHC}.
Theoretical uncertainties were evaluated from variations of factorization and renormalization scales, \alphas, and PDFs. Only LQs coupling to the third-generation quarks and either exclusively to the first-generation leptons or exclusively to the second-generation leptons were considered.
To ensure that LQs decay promptly, the coupling parameter $\lambda$ was set to give a LQ width of about 0.2\% of its mass. \MADSPIN\cite{madspin1,madspin2} was used to decay top quarks while preserving the spin-correlation and finite-width effects.
For this analysis, signal samples were produced for LQ mass values from 900 to 2000~GeV, with a 100~GeV step size in general and a finer 50~GeV step size near the expected LQ mass exclusion limits, and $\beta=1.0$ with fully hadronic top decays.
 
The dominant backgrounds in this search are \zjets{} and \ttbar{} production, with two leptons in the final state. Sources of smaller backgrounds considered include single top quark, $\ttbar V$ ($V$ = $W$, $Z$), diboson ($WZ$, $ZZ$, $WW$), and \wjets{} production. The background contribution from multi-jet production was found to be negligible and is not considered in this search.
 
The \zjets{}, \wjets{} and diboson samples were generated using \SHERPAV{2.2.1}~\cite{Sherpa} with the NNPDF3.0nnlo PDF set. The \zjets{} and \wjets{} samples were normalized to the NNLO cross sections calculated with FEWZ~\cite{Vjet_NNLO}. Matrix elements were calculated for up to two partons at NLO and four partons at leading order (LO) using \COMIX~\cite{Comix} and \OPENLOOPS~\cite{Openloops1,Openloops2,Openloops3} matrix-element generators, and merged with the \SHERPA parton shower~\cite{sherpa-parton-shower} using the ME+PS@NLO prescription~\cite{MEPS1,MEPS2,MEPS3,MEPS4}. For the diboson samples, matrix elements were calculated for up to one parton at NLO and three partons at LO using \COMIX and \OPENLOOPS matrix-element generators, and merged with the \SHERPA parton shower using the ME+PS@NLO prescription.
 
The nominal \ttbar{} and single-top event samples in the $Wt$-, t- and s-channels were simulated with \POWHEGBOXV{v2}~\cite{Powheg1,Powheg2,Powheg3,Powheg4,Powheg_singletop1,Powheg_singletop2} which provides matrix elements at NLO in \alphas with the NNPDF3.0nlo PDF set. The \POWHEGBOX event generator was interfaced with \PYTHIAV{8.230} for the parton shower and hadronization, using the \textsc{A14} tune and the NNPDF2.3lo PDF set. The NLO radiation factor, $h_{\mathrm{damp}}$, was set to 1.5 times the mass of the top quark, $m_{\text{top}}$.
The diagram removal (DR) method was used to remove the interference between $Wt$-channel single-top production and \ttbar{} production~\cite{tW_interference1}. The related uncertainty is estimated by comparison with an alternative sample generated using the diagram subtraction (DS) scheme~\cite{tW_interference1,ATL-PHYS-PUB-2016-020}.
The \ttbar{} samples were normalized to the NNLO cross section with soft-gluon resummation to NNLL accuracy using \TOPplusplusV{2.0}~\cite{ttbar_NNLO1,ttbar_NNLO2,ttbar_NNLO3,ttbar_NNLO4,ttbar_NNLO5,ttbar_NNLO6,ttbar_NNLO7}.
The single-top cross sections for the t- and s-channels are normalized to their NLO predictions using \HATHORV{2.1}~\cite{Hathor1,Hathor2}, while for the $Wt$-channel the cross section is normalized to its NLO+NNLL prediction~\cite{Kidonakis:2010ux,Kidonakis:2013zqa}.
To estimate the modelling uncertainties from the choice of generator and parton shower, alternative samples were generated at NLO for both the \ttbar{} and single-top events using \MGMCatNLOV{2.6.0} interfaced to \PYTHIAV{8.230}, and \POWHEGBOXV{v2} interfaced to \HERWIGV{7.04}~\cite{Herwig1,Herwig2}, respectively.
 
The $\ttbar V$ samples were simulated using \MGMCatNLOV{v2.3.3}~\cite{Madgraph} at NLO in \alphas with the NNPDF3.0nlo PDF set. \MADGRAPH was interfaced with \PYTHIAV{8.210}~\cite{Pythia82} using the A14 tune and NNPDF2.3lo PDF set for parton showering and hadronization.
The cross sections of the samples were calculated at NLO QCD and NLO EW accuracy using \MGMCatNLO as reported in Ref.~\cite{ttV_NLO1}. In the case of $\ttbar \ell\ell$ the cross section is additionally scaled by an off-shell correction estimated at one-loop level in \alphas.
 
The $\ttbar V$ events used \textsc{EvtGen v1.2.0} ~\cite{evtgen} to simulate the modelling of $b$- and $c$-hadron decays, and all other simulated events, except those generated by \SHERPA, used \textsc{EvtGen v1.6.0}.
 
All simulated event samples for the nominal predictions were passed through the ATLAS simulation infrastructure~\cite{SOFT-2010-01}, using the full $\GEANT4$~\cite{Geant4} simulation of the ATLAS detector. The alternative \ttbar{} and single-top generator samples were processed with a fast simulation~\cite{af2} of the ATLAS detector with parameterized showers in the calorimeters. Simulated events were then reconstructed using the same software as used for the data, and overlaid with additional \pp collisions in the same or nearby bunch crossings (pile-up) simulated using the soft QCD processes of the \PYTHIAV{8.186}~\cite{Pythia81} generator with the NNPDF2.3lo PDF set and the \textsc{A3} tune~\cite{ATL-PHYS-PUB-2016-017}. The Monte Carlo samples were reweighted to match the distribution of the number of pile-up interactions to the data.
\section{Analysis object selection}
\label{sec:objects}
 
A set of physics objects (electrons, muons, jets, and missing transverse momentum) are reconstructed using an optimized combination of information from the various subsystems of the ATLAS detector. The reconstructed primary vertex of the event is required to have at least two associated ID tracks with $\pT > 0.5$~\GeV. If more than one primary vertex candidate is reconstructed in an event, the vertex with the largest $\sum \pT^{2}$ of all associated tracks is considered as the hard-scatter vertex.
 
Electron candidates are reconstructed from clusters of energy deposits in the electromagnetic calorimeter associated with a charged-particle track reconstructed in the ID. To ensure that electron candidates originate from the primary vertex, they are required to possess $|d_{0}|/ \sigma_{d_{0}} < 5$ and $|z_{0} \sin\theta| < 0.5 \text{ mm}$, where $d_{0}$ ($z_{0}$) is the transverse (longitudinal) impact parameter relative to the primary vertex and $\sigma_{d_{0}}$ is the uncertainty in $d_{0}$. The electron candidates are required to satisfy the \emph{Tight} likelihood identification criteria for high purity~\cite{ElectronReco}. High-purity candidates must fulfil the \emph{Loose} isolation criteria with fixed cuts on isolation variables to further suppress background contributions from hadrons that are misidentified as electrons~\cite{ElectronReco}. Additionally, the electrons are required to have $\pT > 30$~\GeV\ and pseudorapidity $|\eta| < 2.47$, while excluding those in the barrel--endcap transition region ($1.37 < |\eta| < 1.52$) of the electromagnetic calorimeters.
 
Muon candidates are reconstructed from a combined measurement of tracks in the inner detector and the muon spectrometer. The associated tracks must point to the primary vertex by satisfying $|d_{0}|/ \sigma_{d_{0}} < 3$ and $|z_{0} \sin\theta| < 0.5$~mm. The muon candidates are required to satisfy the \emph{Medium} muon identification selection criteria~\cite{MuonReco} if the leading muon's \pT is below 800~\GeV; otherwise, tighter \emph{High-Pt} muon identification requirements~\cite{MuonReco} are applied to guarantee the best muon resolution and removal of poorly measured tracks in the high-\pT regime. To reject background from muons originating from hadron decays, the \emph{FixedCutTightTrackOnly} track-based isolation criterion is applied, with a wider isolation cone used for $\pT > 50$~\GeV~\cite{MuonReco}. The muons are required to have $\pT > 30$~\GeV\ and $|\eta| < 2.5$.

Small-radius (Small-$R$) jets are reconstructed using the anti-\kt algorithm~\cite{antikt1,Fastjet} with a radius parameter of $R = 0.4$ and with particle-flow objects~\cite{pflow1,pflow2} as inputs. These particle-flow objects are typically either charged-particle tracks that originate from the hard-scatter vertex and are matched to a set of topo-clusters in the calorimeters~\cite{PERF-2014-07}, or the remaining calorimeter energy clusters after the subtraction of calorimeter energy associated with those charged-particle tracks.
Small-$R$ jets are considered if they satisfy $\pT > 25$~\GeV\ and $|\eta| < 2.5$.
For small-$R$ jets with $\pT < 60$~\GeV\ and $|\eta| < 2.4$, a multivariate jet vertex tagger is employed to reduce contamination by jets coming from pile-up~\cite{jvt}.
Small-$R$ jets are only used for the object overlap removal discussed below and for the event kinematic reconstruction discussed in Section~\ref{sec:SR BDT classification}.
 
Large-$R$ jets are reconstructed from topo-clusters of energy deposits in the calorimeters using the anti-\kt algorithm with a radius parameter of $R = 1.0$. To remove contributions from pile-up, the \kt-based trimming algorithm~\cite{Trimming,kt1,kt2,kt3} is employed to recluster jet constituents into subjets with a finer $R$-parameter value of 0.2 and discard subjets with energy less than 5\% of the large-$R$ jet's energy~\cite{PERF-2012-02}. Trimmed large-$R$ jets are required to have $\pT > 200$~\GeV, $|\eta| < 2.0$ and jet mass $m > 50~\GeV$. To identify large-$R$ jets that are likely to have originated from the hadronic decay of a top quark, jet substructure information is exploited as inputs to the BDT model in the muon channel, as discussed in Section~\ref{sec:SR BDT classification}, using the N-subjettiness ratio $\tau_{32}$~\cite{tau32_a,tau32_b}, the splitting measure $\sqrt{d_{23}}$~\cite{d23} and the $Q_{W}$ variables~\cite{Qw}.
 
The missing transverse momentum, $E_\mathrm{T}^\mathrm{miss}$, in a given reconstructed event is computed as the magnitude of the negative vector sum of the \pT of all reconstructed leptons and small-$R$ jets. A track-based soft term is also included in the $E_\mathrm{T}^\mathrm{miss}$ calculation to account for the `soft' energy from inner detector tracks that are not matched to any of the selected objects but are consistent with originating from the primary vertex~\cite{PERF-2016-07,ATLAS-CONF-2018-023}.
 
To avoid double counting of the same object in different reconstructed object types, an overlap removal procedure is applied to specific pairs of objects that either share a track or have small separation in $\Delta R$.
Electron candidates are discarded if they are found to share a track with a more energetic electron or a muon. For overlapping small-$R$ jets and electrons, small-$R$ jets within $\Delta R=0.2$ of a reconstructed electron are removed. If the nearest surviving small-$R$ jet is within $\Delta R=0.4$ of the electron, then the electron is discarded. To reject hadronic jet candidates produced by bremsstrahlung from very energetic muons, the jet is required to have at least three associated tracks if it lies within a cone of $\Delta R=0.2$ around a muon candidate. However, if a surviving jet is separated from the nearest muon with transverse momentum $p^{\mu}_{\mathrm{T}}$ by $\Delta R < 0.04+10\,\GeV/p^{\mu}_{\mathrm{T}}$ up to a maximum of 0.4, the small-$R$ jet is kept and the muon is removed instead; this reduces the background contributions due to muons from hadron decays. No dedicated overlap-removal procedure between large-$R$ and small-$R$ jets is performed. As high-\pT{} electrons could deposit significant amounts of energy in the calorimeter to form large-$R$ jets, the electron energy is removed from any overlapping large-$R$ jets before the jet momentum requirements are applied to avoid double counting the electrons as large-$R$ jets. This approach has a 20\% better signal efficiency compared to rejecting large-$R$ jets that overlap with a reconstructed electron.
\section{Analysis strategy}
\label{sec:analysis}
 
\subsection{Event selection}
In the signal region (SR), events were recorded using either a set of single-electron triggers or a set of single-muon triggers.
The single-electron triggers imposed a \pT threshold of 26~\GeV\ (24~\GeV\ in 2015) and isolation requirements, or a \pT threshold of 60~\GeV\ and no isolation requirements~\cite{TRIG-2018-05}.
The single-muon triggers accepted an isolated muon with $\pT > 26$~\GeV\ (20~\GeV\ in 2015) or any muon with $\pT > 50$~\GeV~\cite{TRIG-2018-01}.
Events with exactly two opposite-sign, same-flavour leptons with \pT above 100~\GeV\ are considered. Events must also have at least two large-$R$ jets.
In addition, events containing a lepton pair with invariant mass below 120~\GeV\ are removed to reduce background contributions from low-mass resonances.
In the SR, the dominant backgrounds are from the \ttbar{} and \zjets{} processes.
The LQ signal, \ttbar{} and \zjets{} events which satisfy these SR criteria are used to train a BDT for signal and background classification.
 
Dedicated control regions (CRs) are defined in order to extract the normalization of the \ttbar{} and \zjets{} backgrounds from data.
For the \ttbar{}-enriched CR, the selection criteria are the same as in the SR, except that either a single-electron trigger or a single-muon trigger must be satisfied and events must contain exactly one opposite-sign electron--muon pair.
The \zjets{}-enriched CR is kept orthogonal to the SR by selecting data in a dilepton invariant mass window $70 < m_{\ell\ell} < 110$~\GeV\ around the $Z$ boson mass.
A summary of the event selections for the signal and control regions is given in Table~\ref{tab:selection_summary}.
 
The expected numbers of events in the SR for the background processes and signal hypothesis with mass $m_{\textrm{LQ}} = 1500$~\GeV\ are shown in Table~\ref{Tab:FinalYields}.
For a signal model with $\beta=1$ and a fully hadronic top-quark final state, the acceptance times efficiency of the SR selection, for LQ masses from $m_{\textrm{LQ}}=900$ to 2000~\GeV, ranges from 32 to 49\% in the electron channel, and from 36 to 43\% in the muon channel.
 
\begin{table}[!h]
\begin{center}
\caption{Summary of event selections applied in the signal and control regions. Leptons and large-$R$ jets are, respectively, denoted by $\ell$ and J.}
\label{tab:selection_summary}
\begin{tabular}{l c c c}
\hline
& \ttbar{} CR & \zjets{} CR & SR \\
\toprule
Leptons & \multicolumn{3}{c}{$p_{\mathrm{T}}^{\mathrm{\ell}} > 100~\GeV$, $|\eta_{e}| < 2.47$, $|\eta_{\mu}| < 2.5$}\\
& \multicolumn{3}{c}{$N_{\ell} = 2$; opposite-sign}\\
Large-$R$ jets & \multicolumn{3}{c}{$p_{\mathrm{T}}^{\mathrm{J}} > 200~\GeV$, $|\eta_{\mathrm{J}}| < 2.0$, $m_{\mathrm{J}} > 50~\GeV$}\\
& \multicolumn{3}{c}{$N_{\mathrm{J}} \geq 2$}\\
Dilepton invariant mass\quad & $m_{\ell\ell}$ > 120~\GeV & 70~\GeV\ < $m_{\ell\ell}$ < 110~\GeV & $m_{\ell\ell}$ > 120~\GeV \\
Lepton flavour & $e\mu$ & \multicolumn{2}{c}{$ee\:$or$\:\mu\mu$}\\
\hline
\end{tabular}
\end{center}
\end{table}
 
\subsection{Signal region BDT classification}
\label{sec:SR BDT classification}
A BDT classifier is trained in the SR to further separate the signal from the backgrounds.
A gradient boosting approach is used with the \XGBOOST framework~\cite{xgboost} as the back end for mathematical computations.
 
The gradient boosting algorithm contains at most 1000 trees with a maximal tree depth of 3, while early stopping is employed if no improvement in the classification is found after 10 iterations of the trees.
To avoid overtraining the classifier, nested cross validation~\cite{Stone1976CrossValidatoryCA} was performed to obtain an unbiased evaluation of the classifier performance.
The classifier produces an output score referring to the predicted probability that the event contains LQs, which is then used as the final discriminant to separate LQ signal events from the SM backgrounds.
 
A natural basis of kinematic observables can be created, utilizing Lorentz symmetry to reduce unnecessary duplication of observables, in the rest frames of intermediate particle states, conditioned on the hypotheses of LQ pair, dileptonic \ttbar{} or \zjets{} decay processes.
A suite of such discriminating variables is constructed using the recursive jigsaw reconstruction technique~\cite{RJR}, and is provided as inputs to the classifier.
The dileptonic \ttbar{} reconstruction scheme is based on the `min $\Delta M_{\mathrm{top}}$ approach' of the recursive jigsaw reconstruction technique, in which the two leading small-$R$ jets are used as the $b$-quark candidates from the top-quark decays.
Variables related to hadronic and leptonic activity, missing transverse momentum and jet substructure are also used to provide additional separation power.
Large-$R$ jet substructure variables are only used in the muon channel. As discussed in Sect.~\ref{sec:objects}, the energy of electrons overlapping with large-$R$ jets is subtracted from the jet four-momentum to avoid double counting. Such kinematic modification of large-$R$ jets is incompatible with the use of substructure variables.
In total, 29 inputs are used in the BDT classifier in the electron channel and 32 in the muon channel. The top five discriminating variables are the dilepton invariant mass, the scalar \pT sum of the two leptons, the two large-$R$ jet masses, and the reconstructed LQ mass. Fig.~\ref{CRplots_inputs} shows the distributions of the dilepton invariant mass in the \zjets{} CR of the muon channel, and the reconstructed $W$ mass based on a dileptonic \ttbar{} hypothesis in the \ttbar{} CR of the electron channel. In general, the kinematic variables show good agreement between data and the background expectation in the CRs.
A complete list of the input variables is provided in Table~\ref{tab:discriminating_variables}.
 
\begin{figure}[htb!]
\centering
\includegraphics[width=0.48\textwidth]{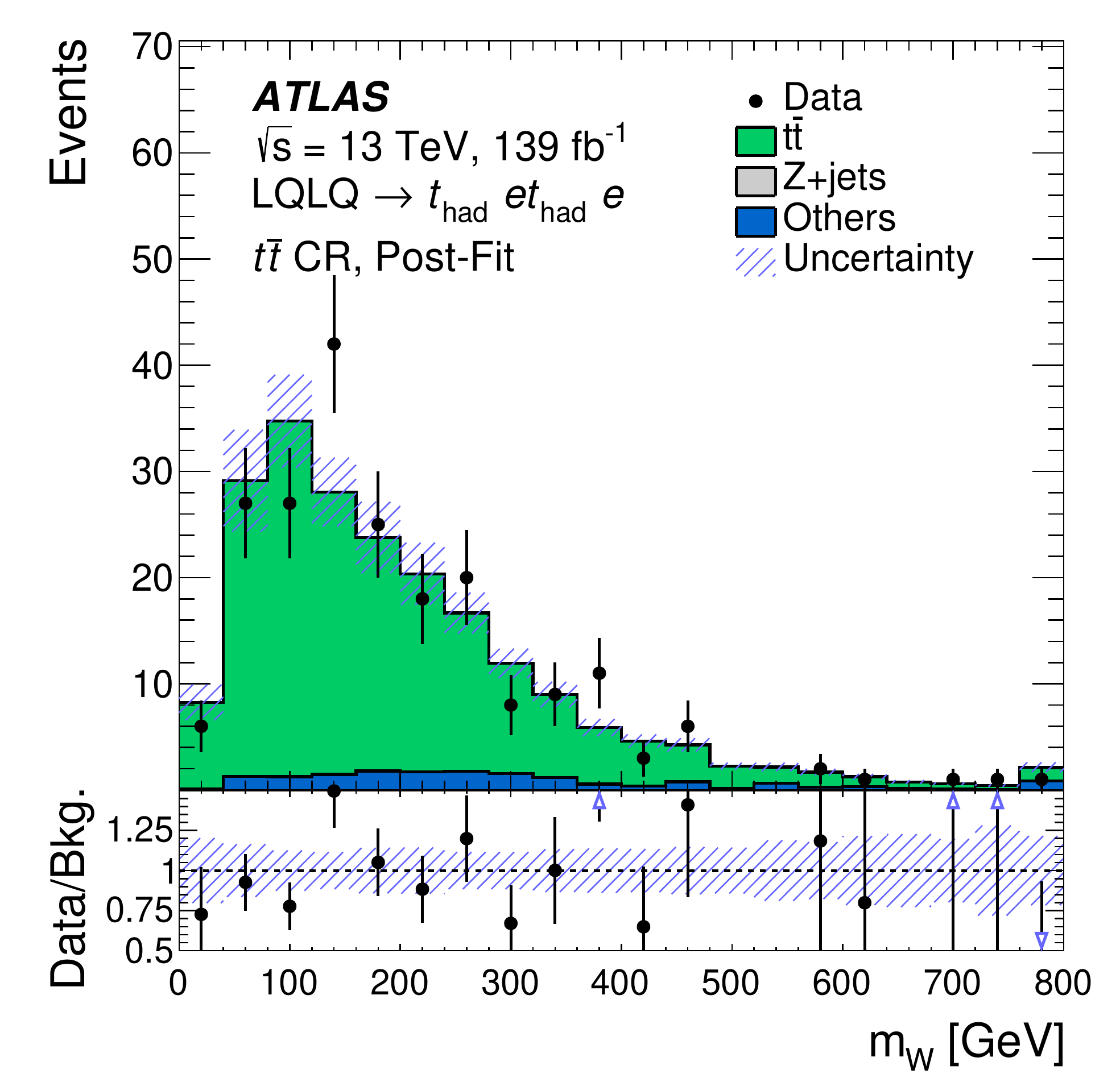}
\includegraphics[width=0.48\textwidth]{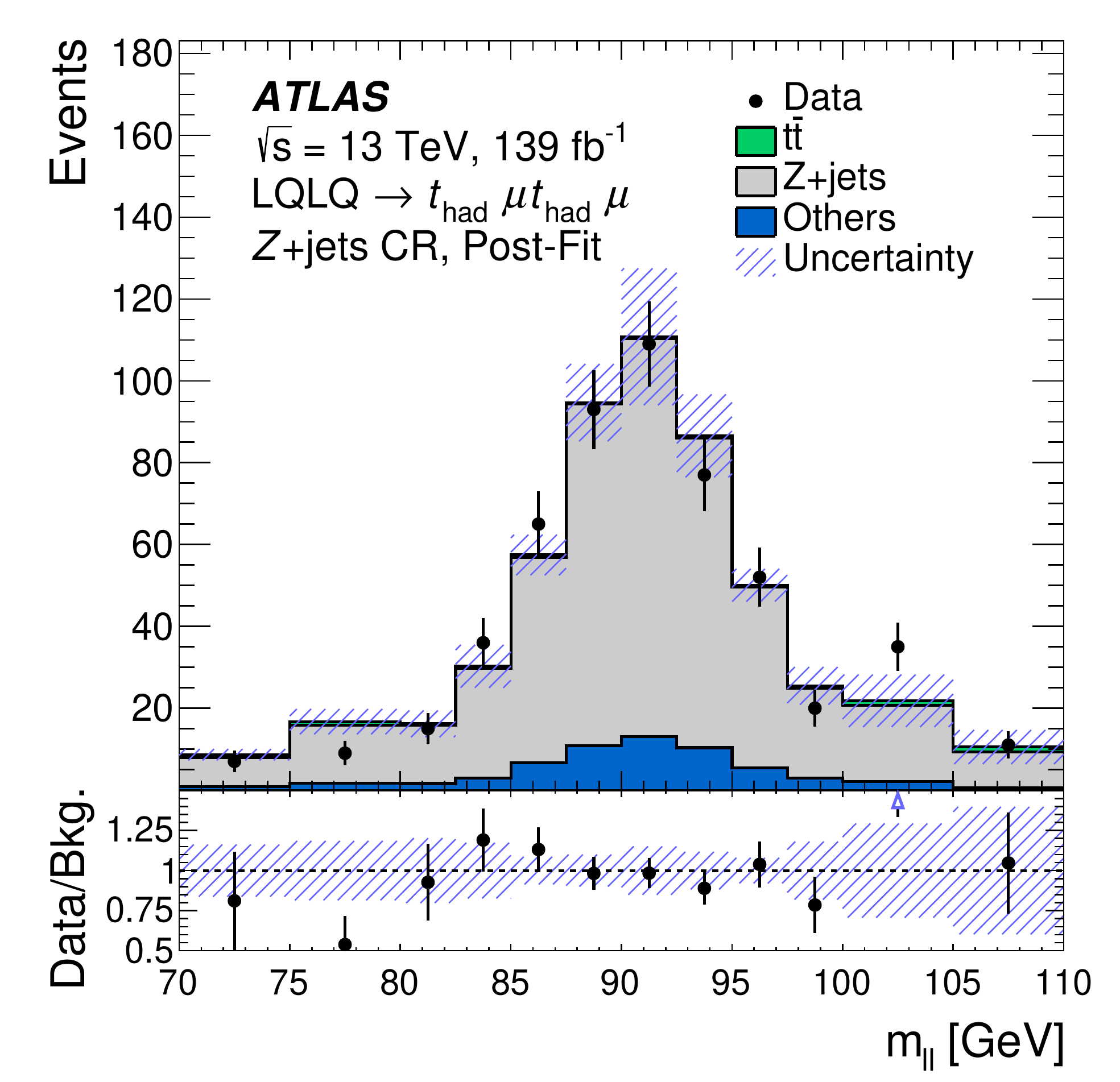}
\caption{Distributions of the reconstructed $W$ mass associated with the leading lepton assuming a dileptonic hypothesis in the \ttbar{} CR after the simultaneous background-only fit of the electron channel CRs (left), and the dilepton invariant mass $m_{\ell \ell}$ in the \zjets{} CR after the simultaneous background-only fit of the muon channel CRs (right). The bottom panels show the ratio of data to expected background. The hatched band represents the total uncertainty. The blue triangles indicate points that are outside the vertical range.}
\label{CRplots_inputs}
\end{figure}
 
In order to maximize the sensitivity of the BDT over a wide mass range, and ensure a smooth interpolation of the signal efficiency between the mass points where it was trained, a parameterized machine-learning approach~\cite{parameterization} is implemented.
The inputs to the BDT classifier are expanded to include the theoretical LQ mass, resulting in a single BDT classifier that smoothly provides optimized discrimination across a range of masses, from 900 to 2000~\GeV.
The parameterized BDT was trained with large samples of simulated signal events at $m_{\textrm{LQ}}$ values from 900 to 1900~\GeV, with a 200~\GeV\ step size.
The modelling of the BDT distribution of the main backgrounds is validated using events in the \zjets{} and \ttbar{} control regions, as shown in Fig.~\ref{CRplots_outputs_1500GeV}.
The number of bins and their boundaries in the SR are optimized to maximize the expected scalar leptoquark sensitivity while ensuring a minimum of three background events in the highest BDT bin. It was found that having three bins in the SR was optimal, which are defined as the low, mid and high BDT SR. Of the signal events which enter the signal region, over 94\% fall into the high BDT SR while only 1\% and 8\% of the \ttbar{} and \zjets{} background do so, respectively.
 
\begin{figure}[htb!]
\centering
\includegraphics[width=0.48\textwidth]{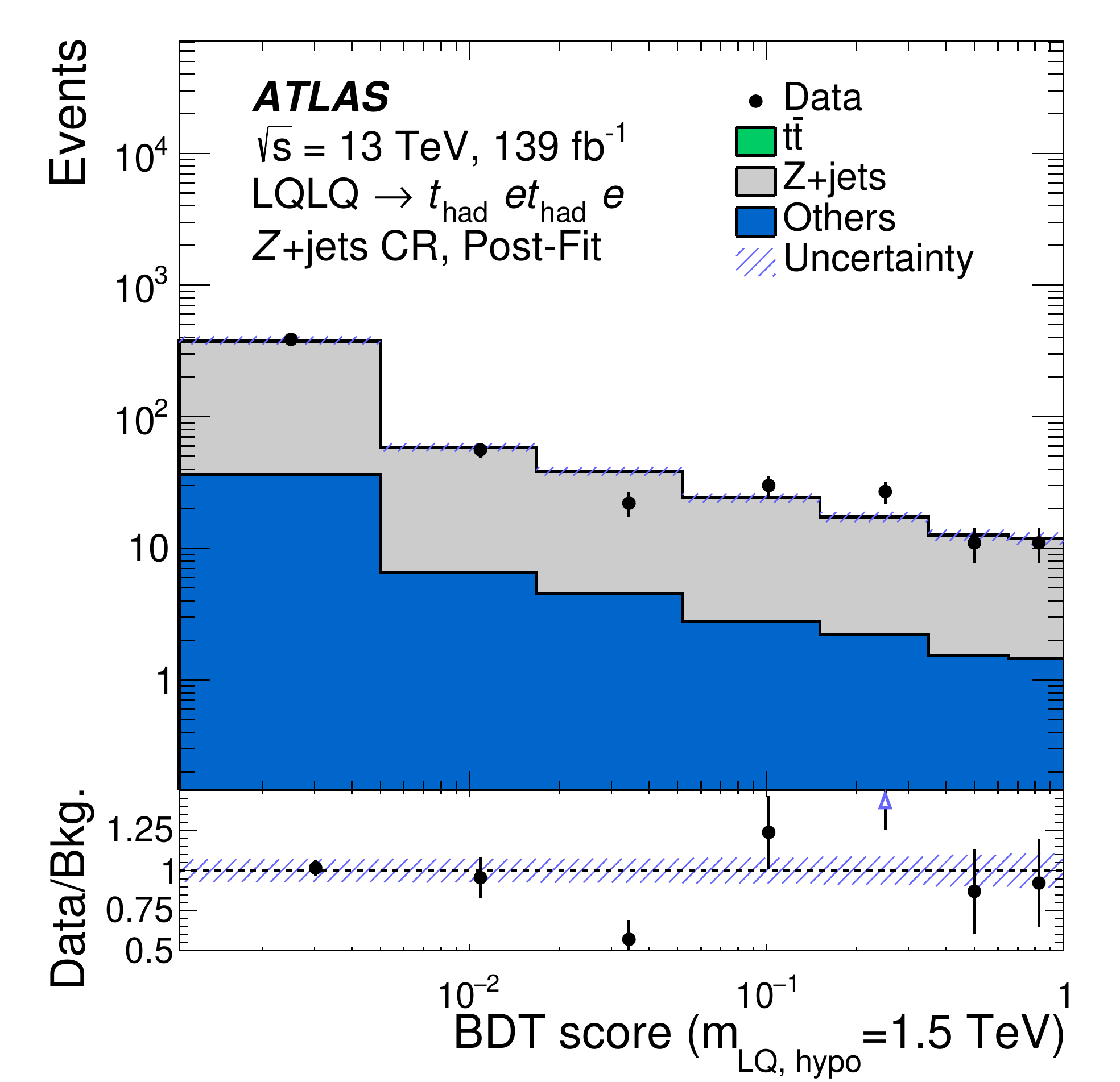}
\includegraphics[width=0.48\textwidth]{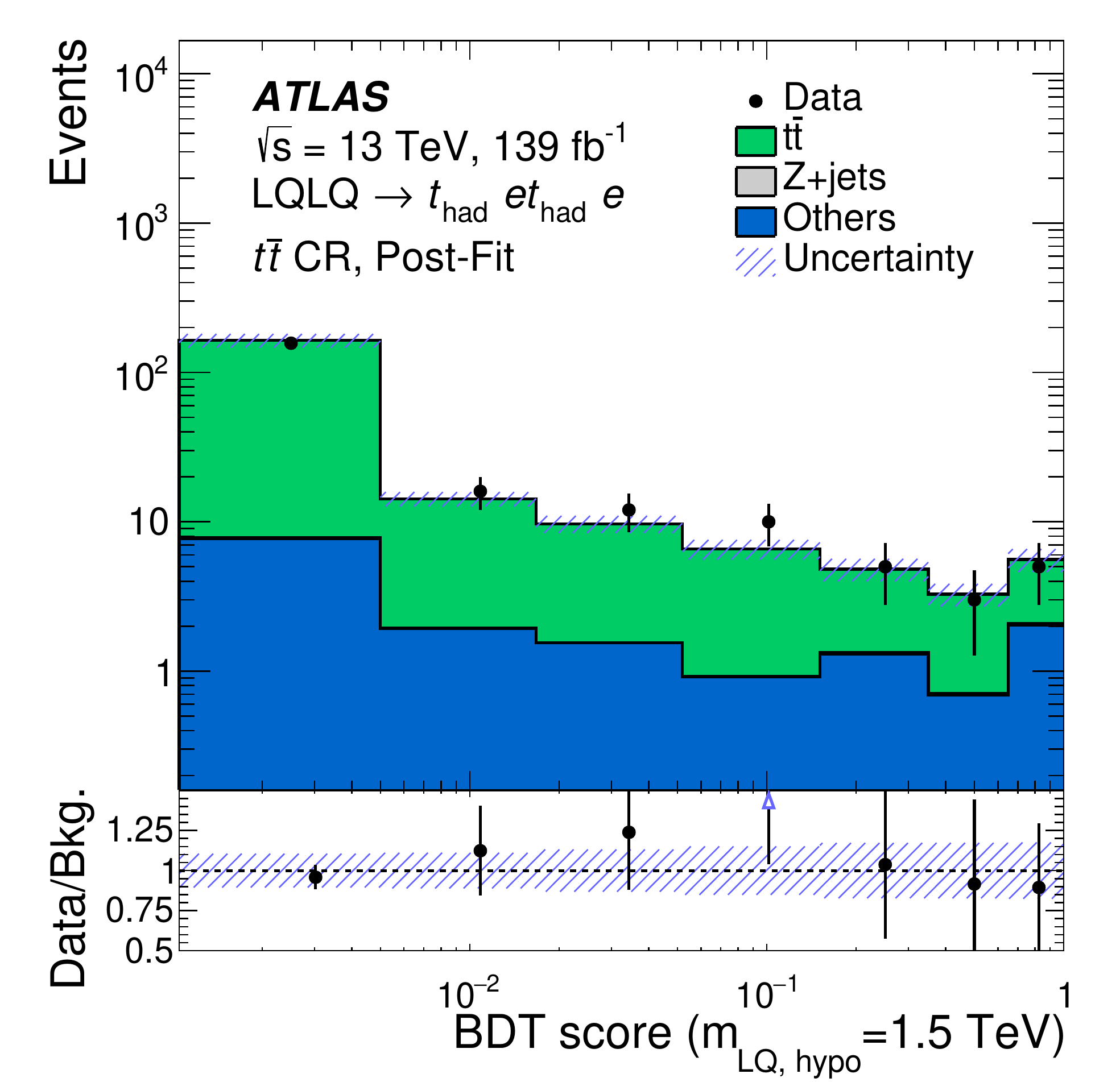}
\includegraphics[width=0.48\textwidth]{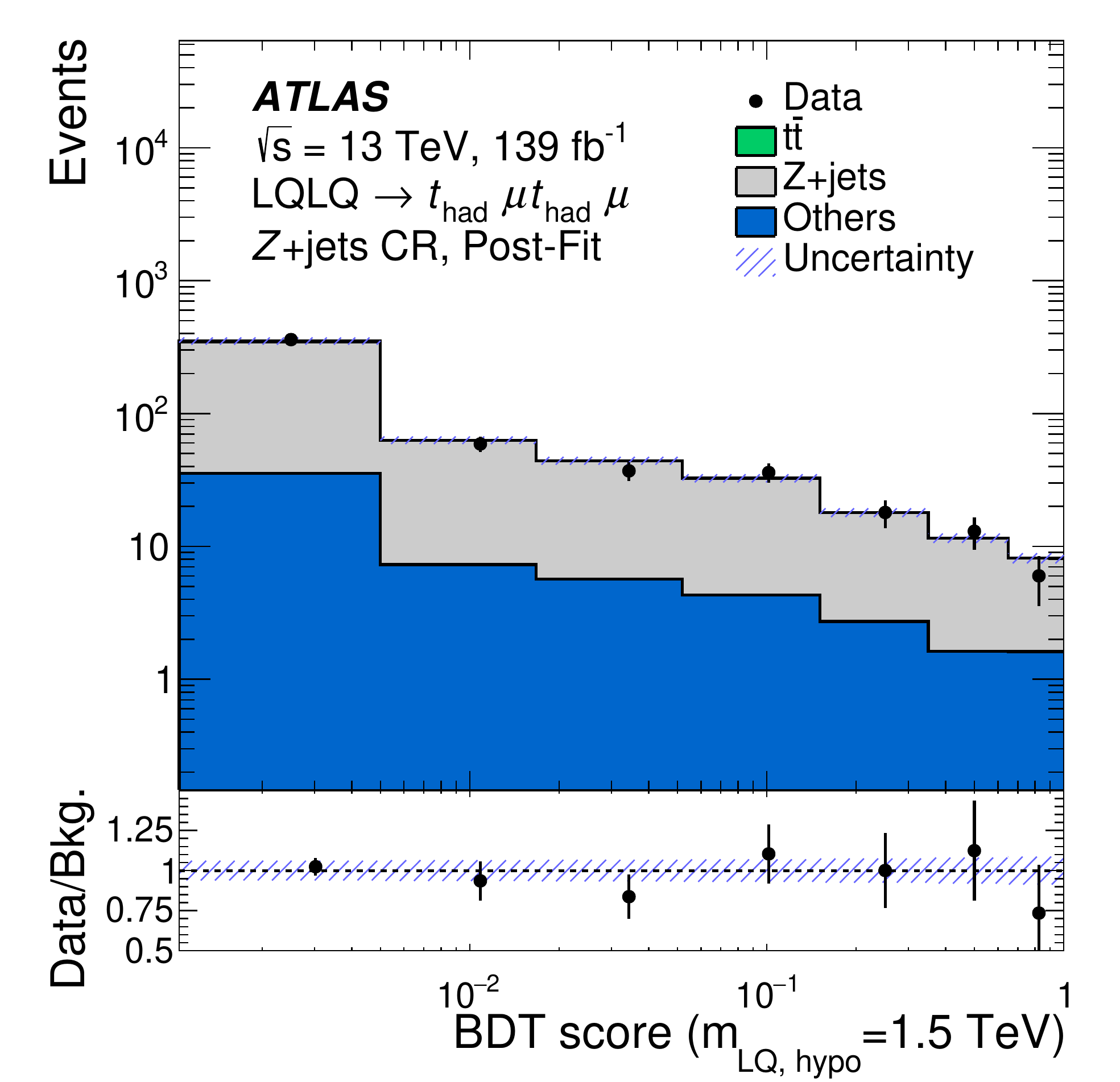}
\includegraphics[width=0.48\textwidth]{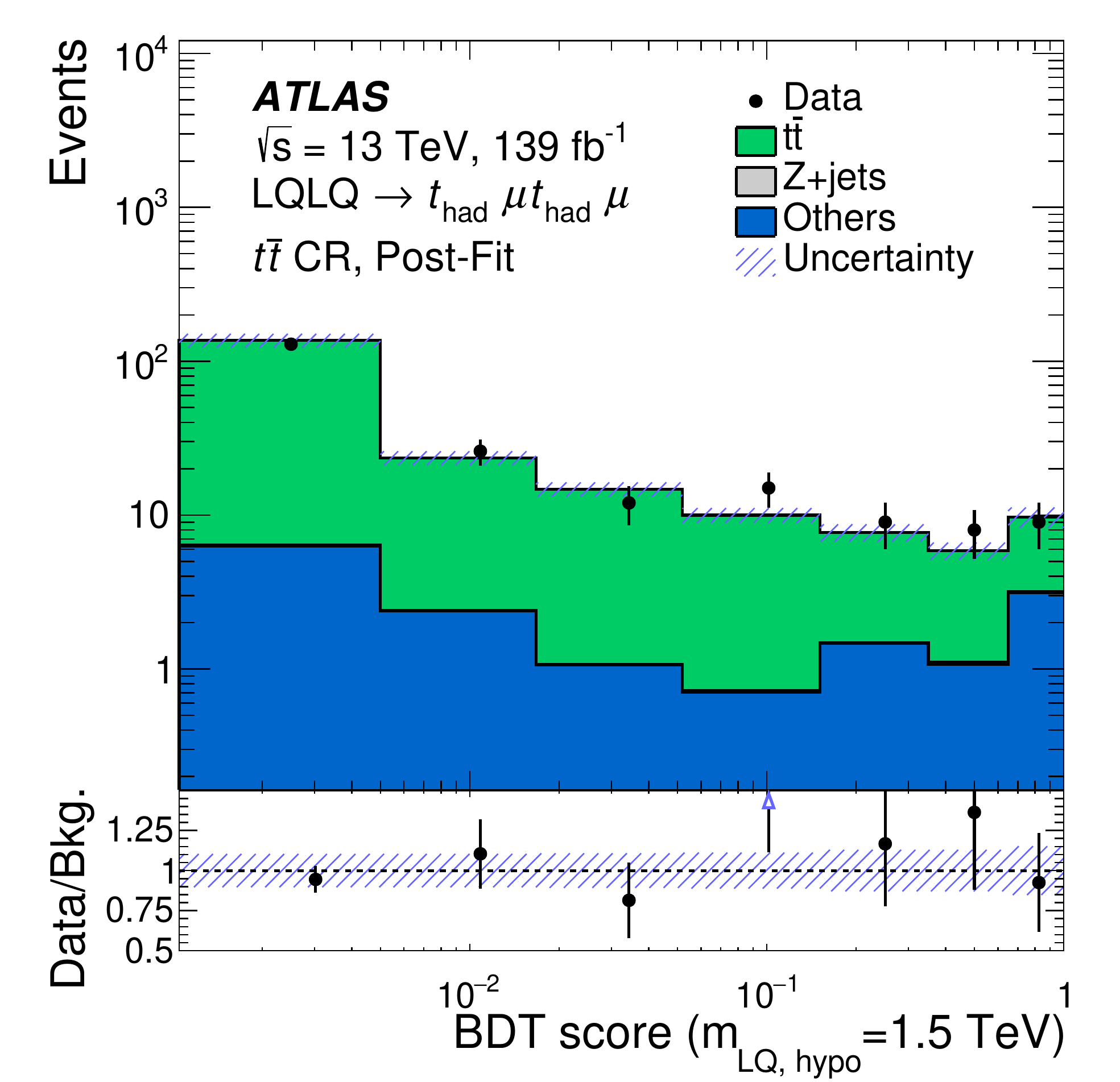}
\caption{Distributions of the BDT output score in the \zjets{} and \ttbar{} CRs for the electron (top row) and muon (bottom row) channel after the simultaneous background-only fit of the CRs. The bottom panels show the ratio of data to expected background. The hatched band represents the total uncertainty. The blue triangles indicate points that are outside the vertical range. All BDT scores correspond to the theoretical LQ mass parameter $m_{\textrm{LQ, hypo}}$ set to 1.5~\TeV. The first bin contains all underflow events.}
\label{CRplots_outputs_1500GeV}
\end{figure}
\section{Systematic uncertainties}
\label{sec:systematics}

The systematic uncertainties are broken down into three broad categories: luminosity and cross-section uncertainties,
detector-related experimental uncertainties, and modelling uncertainties in simulated background processes.
The uncertainty from each source is treated as a Gaussian-distributed or log-normal nuisance parameter in a profile-likelihood fit of the CR normalizations and BDT output score distributions, and shape effects are taken into account where relevant.
Due to the tight selection criteria applied and resultant statistical limitation to the sensitivity, the systematic uncertainties only mildly degrade the sensitivity of the search.
 
\subsection{Luminosity and normalization uncertainties}
\label{sec:systematic_Luminosity}
 
The uncertainty in the combined 2015--2018 integrated luminosity is 1.7\%~\cite{ATLAS-CONF-2019-021}, obtained using the LUCID-2 detector~\cite{LUCID2} for the primary luminosity measurements. 
 
Theoretical cross-section uncertainties are applied to the various simulated samples.
For the LQ signal, PDF, \alphas and scale uncertainties are considered in the approximate NNLO + NNLL calculation of the cross section. The PDF and \alphas uncertainties are estimated from the PDF4LHC15 error set~\cite{pdf4lhc}. The effect of uncertainties in the renormalization and factorization scales is estimated from variations by a factor of two about the central scales.
The overall uncertainty ranges from 10\% at low LQ masses to 25\% at 2~\TeV~\cite{stop-xs1,stop-xs2,stop-xs3,stop-xs4}.
This cross-section uncertainty is not included in the profile-likelihood fit, but represented by an uncertainty band around the theoretical prediction in the cross-section limit plots in Section~\ref{sec:result}.
The uncertainties for $W$+jets and diboson production are both assumed to be 50\%~\cite{Vjets_xs, ATL-PHYS-PUB-2016-002}.
For single top quark and $\ttbar V$ production, the uncertainties are taken as 7\%~\cite{Hathor1,Hathor2} and 30\%~\cite{ATL-PHYS-PUB-2016-005}, respectively.
The normalizations of \ttbar{} and \zjets{} are determined from data via unconstrained normalization parameters.
 
\subsection{Detector-related uncertainties}
\label{sec:systematic_detector}
The dominant sources of detector-related uncertainties in the signal and background yields relate to the lepton identification efficiency scale factors that are used to correct for the difference between the Monte Carlo simulation and data.
These uncertainties have an impact on the fitted signal yield of roughly 12\% and 5\% in the electron and muon channel respectively.
Additional uncertainties to account for the degradation of the muon momentum resolution due to the impact of possible misalignment between layers of the MS, as well as between the MS and the ID, were estimated to be 5\%.
 
Uncertainties in the small-$R$ and large-$R$ jet energy scales and resolutions are also considered. The small-$R$ and large-$R$ jet energy scales and their uncertainties are derived by combining information from test-beam data, LHC collision data and simulation~\cite{ATL-PHYS-PUB-2015-015}.
The uncertainties in the jet energy scale have an impact of up to $\sim$4\% on the fitted signal yield.
Moreover, in the case where an electron overlaps with a large-$R$ jet, the impact on the jet energy scale calibration due to the analysis-specific removal of the electron energy from the large-$R$ jets was evaluated.
The jet axis shift and the fraction of calibrated jet energy contributed by the overlapping electrons were studied in simulated events.
These additional jet systematic uncertainties have an impact of $<$3\% on the signal yield.
 
Other detector-related uncertainties come from uncertainties in the large-$R$ jet mass scales and resolutions; lepton isolation and reconstruction; lepton trigger efficiencies, energy scales, and resolutions; the \met{} reconstruction; pile-up modelling; and the jet-vertex-tagger requirement. Uncertainties in the object momenta are propagated to the \met{} measurement, and additional uncertainties in \met{} arising from the `soft' energy are also considered. These all have negligible impact on the fitted signal yield ($<$3\% each).

\subsection{Generator modelling uncertainties}
\label{sec:systematic_modelling}
 
Modelling uncertainties are estimated for the signal as well as $Z$+jets, \ttbar{} and single-top-quark backgrounds.
The modelling uncertainties are estimated by comparing simulated samples generated with different configurations, described in Sect.~\ref{sec:samples}.
 
For the LQ signal, in addition to the cross-section uncertainties, the impact on the acceptance due to variations of the QCD scales, PDF and shower parameters was studied. These uncertainties were estimated from the envelope of independent pairs of renormalization and factorization scale variations by a factor of 0.5 and 2, by propagating the PDF and \alphas uncertainties following the PDF4LHC15 prescription, and by considering two alternative samples generated with settings that increase or decrease the amount of QCD radiation.
Both the PDF and scale variations have an impact below 15\% for all bins considered, while variations of the underlying-event modelling have only a 1--2\% effect.
 
For the $Z$+jets backgrounds, scale, PDF and \alphas variations are considered and their effects are evaluated within the \SHERPA event generator.
Seven variations are considered for the renormalization and factorization scales, with the maximum shift within the envelope of those variations taken to estimate the effect of the scale uncertainty.
The PDF variations include the variation of the nominal NNPDF3.0nnlo PDF as well as the central values of two other PDF sets, MMHT2014nnlo68cl~\cite{mmht} and CT14nnlo~\cite{ct14}. The intra-PDF uncertainty is estimated as the standard deviation of the 100 variations of the NNPDF3.0nnlo set. The envelope of the differences between the nominal and alternative PDF sets is used as an additional nuisance parameter.
The effect of varying \alphas from its nominal value of 0.118 by $\pm0.001$ is also considered. The dominant effect is from the renormalization and factorization scale variations and is about 6\% of the signal yield.
 
For the \ttbar{} background, four sources of modelling uncertainties are considered.
The uncertainty in the matrix-element calculation is estimated by comparing events generated with two different Monte Carlo generators, \MGMCatNLO\ and \POWHEGBOX, while keeping the same parton shower model.
The uncertainty in the fragmentation, hadronization and underlying-event modelling is estimated by comparing two different parton shower models, \PYTHIA\ and \HERWIG, while keeping the same hard-scatter matrix-element calculation.
The effects of extra initial- and final-state gluon radiation are estimated by comparing simulated samples generated with enhanced or reduced initial-state radiation, doubling the $h_\text{damp}$ parameter, and using different values of the radiation parameters~\cite{ATL-PHYS-PUB-2016-020}.
The PDF uncertainty is estimated from the PDF4LHC15 error set. The dominant effect is from the final-state radiation estimation uncertainty and is about 6\% of the signal yield.
 
In this analysis, the single-top-quark background comes mainly from the $Wt$-channel and is a minor background. Similarly to \ttbar{}, uncertainties in the hard-scatter generation, the fragmentation and hadronization, the amount of additional radiation, and the PDF are considered. In addition, the uncertainty due to the treatment of the overlap between $Wt$-channel single top quark production and \ttbar{} production is considered by comparing samples using the DS and DR methods (see Sect.~\ref{sec:samples}). The dominant effect is from the uncertainty in the fragmentation and hadronization and is about 7\% of the signal yield.

\section{Statistical interpretation}
\label{sec:stat_ana}

The binned distributions of the BDT score in the SR and the overall number of events in the \ttbar{} and \zjets{} CR are used to test for the presence of a signal. Hypothesis testing is performed using a modified frequentist method as implemented in \textsc{RooStats}~\cite{RooFit,RooFitManual} and is based on a profile likelihood that takes into account the systematic uncertainties as nuisance parameters that are fitted to the data.
A simultaneous fit is performed in the SR and the two CRs, but done separately for the electron and muon channel.
As the \ttbar{} CR is built requiring an electron and muon, the same events are considered in the independent electron and muon channel fits.
 
The statistical analysis is based on a binned likelihood function $\mathcal{L}(\mu,\theta)$ constructed as a product of Poisson probability terms over all bins considered in the search. This function depends on the signal strength parameter $\mu$, a multiplicative factor applied to the theoretical signal production cross section, and $\theta$, a set of nuisance parameters that encode the effect of systematic uncertainties in the signal and background expectations and are implemented in the likelihood function as Gaussian and log-normal constraints.
Uncertainties in each bin due to the finite size of the simulated samples are also taken into account via dedicated constrained fit parameters.
There are enough events in the CRs and the lowest BDT bin in the SR, where the signal contribution is small, to obtain a data-driven estimate of the \ttbar{} and \zjets{} normalizations and hence the normalizations of those two backgrounds are included as unconstrained nuisance parameters, $\mu_{\ttbar}$ and $\mu_{Z}$.
Nuisance parameters representing systematic uncertainties are only included in the likelihood if either of the following conditions are met: the overall impact on the normalization in a given region is larger than 3\%, or any single bin within the region has at least a 3\% uncertainty. This is done separately for each region and for each template (signal or background).
When the bin-by-bin statistical variation of a given uncertainty is significant, a smoothing algorithm is applied.

The test statistic $q_{\mu}$ is defined as the profile likelihood ratio,
$q_{\mu}=-2\mathrm{ln}(\mathcal{L}(\mu,\hat{\hat{\theta}}_{\mu})/ \mathcal{L}(\hat{\mu}, \hat{\theta}))$,
where $\hat{\mu}$ and $\hat{\theta}$ are the values of the parameters that maximize the likelihood function, and $\hat{\hat{\theta}}_{\mu}$ are the values of the nuisance parameters that maximize the likelihood function for a given value of $\mu$. The compatibility of the observed data with the background-only hypothesis is tested by setting $\mu=0$ in the profile likelihood ratio: $q_{0}=-2\mathrm{ln}(\mathcal{L}(0,{\hat{\hat{\theta}}}_{0})/\mathcal{L}({\hat{\mu}},{\hat{\theta}}))$.
Upper limits on the signal production cross section for each of the signal scenarios considered are derived by using $q_{\mu}$ in the so-called CL$_{\mathrm{s}}$ method~\cite{cls_2, cls}.
For a given signal scenario, values of the production cross section (parameterized by $\mu$) yielding CL$_{\mathrm{s}}<0.05$, where CL$_{\mathrm{s}}$ is computed using the asymptotic approximation~\cite{cls_3}, are excluded at $\geq 95$\% confidence level (CL).
 
 
\section{Results}
\label{sec:result}

\subsection{Likelihood fit results}\label{Results:fitResults}
 
\begin{table}[t!]{}
\begin{center}
\caption{Event yields in the signal and control regions before and after the background-only fit to data in the electron and muon channel. The quoted uncertainties include statistical and systematic uncertainties; for the $t\bar{t}$ and $Z\mathrm{\,+\,jets}$ backgrounds no cross-section uncertainty is included since it is a free parameter of the fit. The contributions from single top, $t\bar{t}V$, diboson and $W\mathrm{\,+\,jets}$ production are included in the `Others' category. In the post-fit case, the uncertainties in the individual background components can be larger than the uncertainty in the sum of the backgrounds, due to the correlations between the fit parameters. Both signal models correspond to $m_{\textrm{LQ}} =$ 1500 GeV assuming 100\% branching ratio into a hadronically decaying top quark and a charged lepton.}
\label{Tab:FinalYields}
\bigskip
\resizebox{0.95\textwidth}{!}{
\centering
\begin{tabular}{ l r@{$\;\pm\;$}l
r@{$\;\pm\;$}l
r@{$\;\pm\;$}l
r@{$\;\pm\;$}l
r@{$\;\pm\;$}l}
\hline
\multicolumn{1}{l}{Sample} & \multicolumn{2}{l}{$t\bar{t}$ CR} & \multicolumn{2}{l}{$Z\mathrm{\,+\,jets}$ CR} & \multicolumn{2}{l}{SR: low BDT} & \multicolumn{2}{l}{SR: mid BDT} & \multicolumn{2}{l}{SR: high BDT} \\
\cmidrule(rl){2-11}
\multicolumn{1}{l}{Electron Channel} & \multicolumn{10}{l}{Pre-fit} \\
\toprule
\,\, $t\bar{t}$   & 222 & 58 & 9.6 & 7.8 & 90 & 30 & 4.3 & 1.9 & 0.6 & 0.3 \\
\,\, $Z\mathrm{\,+\,jets}$   & 0.3 & 0.1 & 520 & 100 & 32.7 & 5.9 & 8.2 & 1.8 & 2.9 & 0.8 \\
\,\, Others   & 16.1 & 5.3 & 55 & 18 & 6.7 & 3.6 & 2.1 & 1.1 & 0.3 & 0.1 \\
\,\, Total background  & 238 & 60 & 590 & 110 & 130 & 36 & 14.6 & 3.4 & 3.7 & 0.9 \\
\,\, Signal ($m_{\textrm{LQ}}$ = 1500 GeV)   & \multicolumn{2}{c}{$<$\,0.001} & 0.006 & 0.002 & \multicolumn{2}{c}{$<$\,0.001} & 0.015 & 0.004 & 7.4 & 1.6 \\
\midrule
\multicolumn{1}{l}{} & \multicolumn{10}{l}{Post-fit} \\
\toprule
\,\, $t\bar{t}$   & 200 & 19 & 10.3 & 5.3 & 86 & 10 & 4.4 & 1.0 & 0.6 & 0.1 \\
\,\, $Z\mathrm{\,+\,jets}$   & 0.22 & 0.04 & 493 & 43 & 30.7 & 2.9 & 8.0 & 0.9 & 2.8 & 0.3 \\
\,\, Others   & 19.1 & 5.7 & 53 & 19 & 9.6 & 5.2 & 3.1 & 1.6 & 0.3 & 0.1 \\
\,\, Total background & 219 & 18 & 556 & 38 & 126 & 12 & 15.4 & 2.0 & 3.7 & 0.3 \\
\,\, Data   & \multicolumn{2}{c}{208} & \multicolumn{2}{c}{544} & \multicolumn{2}{c}{130} & \multicolumn{2}{c}{22} & \multicolumn{2}{c}{6} \\
\hline
\multicolumn{1}{l}{Muon Channel} & \multicolumn{10}{l}{Pre-fit} \\
\toprule
\,\, $t\bar{t}$   & 222 & 58 & 8.9 & 6.9 & 112 & 23 & 8.3 & 5.0 & 0.8 & 0.5 \\
\,\, $Z\mathrm{\,+\,jets}$   & 0.3 & 0.1 & 532 & 45 & 31.7 & 2.8 & 11.7 & 1.3 & 2.9 & 0.3 \\
\,\, Others   & 16.1 & 6.9 & 59 & 19 & 7.6 & 4.1 & 2.2 & 1.7 & 0.6 & 0.4 \\
\,\, Total background  & 238 & 60 & 600 & 53 & 152 & 24 & 22.2 & 6.2 & 4.2 & 1.0 \\
\,\, Signal ($m_{\textrm{LQ}}$ = 1500 GeV)   & \multicolumn{2}{c}{$<$\,0.001} & 0.013 & 0.003 & \multicolumn{2}{c}{$<$\,0.001} & 0.031 & 0.007 & 7.0 & 1.4 \\
\midrule
\multicolumn{1}{l}{} & \multicolumn{10}{l}{Post-fit} \\
\toprule
\,\, $t\bar{t}$   & 187 & 19 & 7.9 & 4.1 & 92.2 & 9.3 & 7.6 & 2.9 & 0.7 & 0.3 \\
\,\, $Z\mathrm{\,+\,jets}$   & 0.22 & 0.03 & 463 & 36 & 27.6 & 2.2 & 10.2 & 1.0 & 2.5 & 0.3 \\
\,\, Others   & 17.9 & 7.5 & 59 & 18 & 8.1 & 4.1 & 2.5 & 1.8 & 0.6 & 0.5 \\
\,\, Total background  & 205 & 19 & 530 & 32 & 127.9 & 9.3 & 20.4 & 3.1 & 3.8 & 0.5 \\
\,\, Data   & \multicolumn{2}{c}{208} & \multicolumn{2}{c}{529} & \multicolumn{2}{c}{123} & \multicolumn{2}{c}{20} & \multicolumn{2}{c}{6} \\
\hline
\end{tabular}
}
\end{center}
\end{table}

The expected and observed event yields in the signal and control regions before and after fitting the background-only hypothesis to data, including all uncertainties, are listed in Table~\ref{Tab:FinalYields}. The total uncertainty shown in the table is the uncertainty obtained from the full fit, and is therefore not identical to the sum in quadrature of each component, due to the correlations between the fit parameters.
A comparison of the post-fit agreement between data and prediction for the signal and control regions is shown in Figure~\ref{FigPostFitResult_BONLY}.
In the electron (muon) channel, the ratio of the \ttbar{} total post-fit yield over the pre-fit yield is $0.90\pm0.25$ ($0.84\pm0.24$).
The ratio of the \zjets{} total post-fit yield over the pre-fit yield is $0.95\pm0.20$ ($0.87\pm0.10$). None of the individual uncertainties are significantly constrained by data.
 
The probability that the data is compatible with the background-only hypothesis is estimated by integrating the distribution of the test statistic, approximated using the asymptotic formulae, above the observed value of $q_{0}$.\footnote{Cross-checks with sampling distributions generated using pseudo-experiments were performed to test the accuracy of the asymptotic approximation for the whole probed leptoquark mass spectrum. The approximation is found to lead to limits that are slightly stronger than those obtained with pseudo-experiments, up to 10\% in general for both channels. The impact of this approximation on the mass limits is below 5~\GeV.}
This value is computed for each signal scenario considered, defined by the assumed mass of the leptoquark.
The lowest local $p$-value is found to be $\sim$11\% (10\%), for a LQ mass of 1450~(1600)~\GeV\ in the electron (muon) channel.
Thus no significant excess above the background expectation is found.

\begin{figure}[htb!]
\centering
\includegraphics[width=0.48\textwidth]{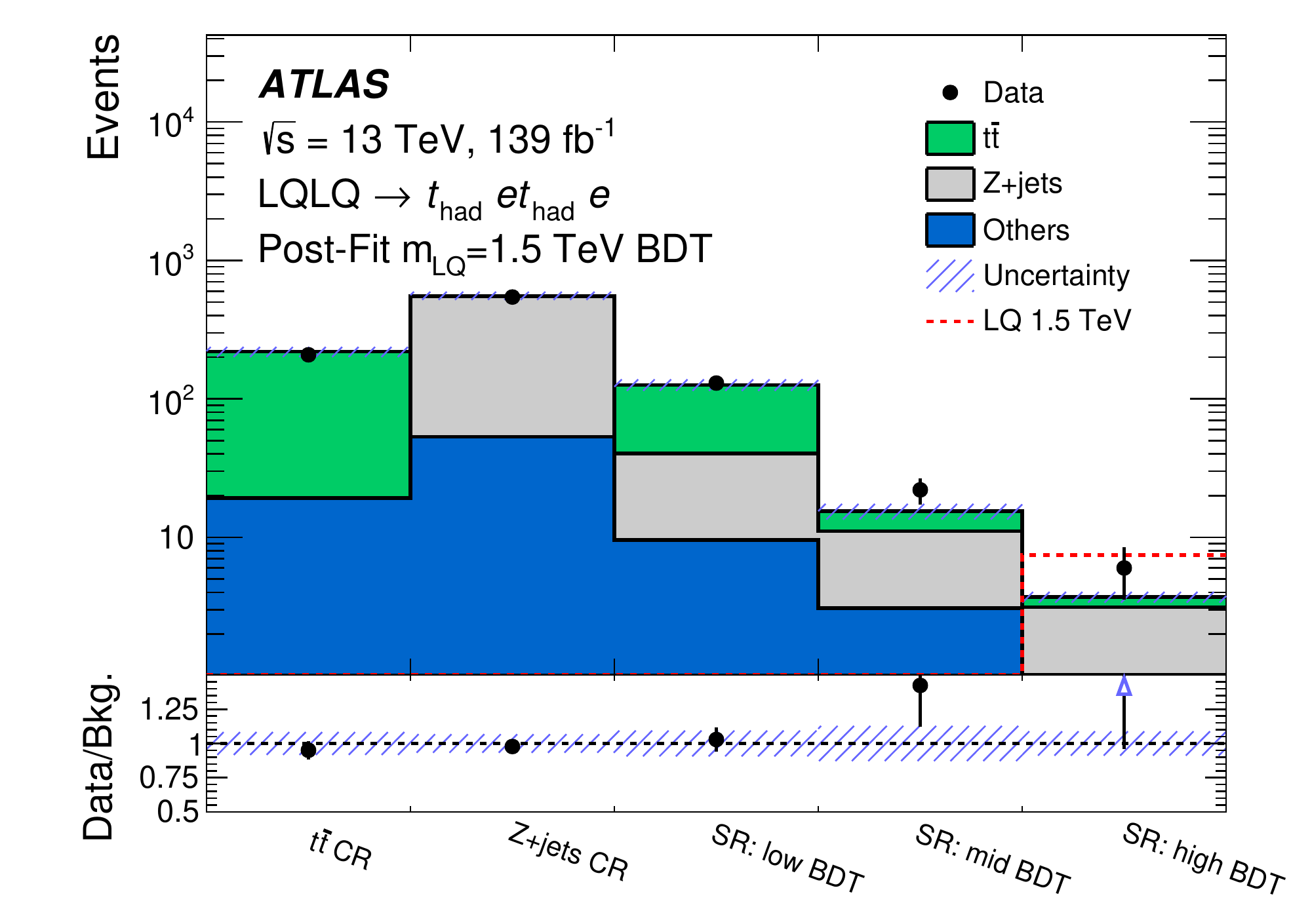}
\includegraphics[width=0.48\textwidth]{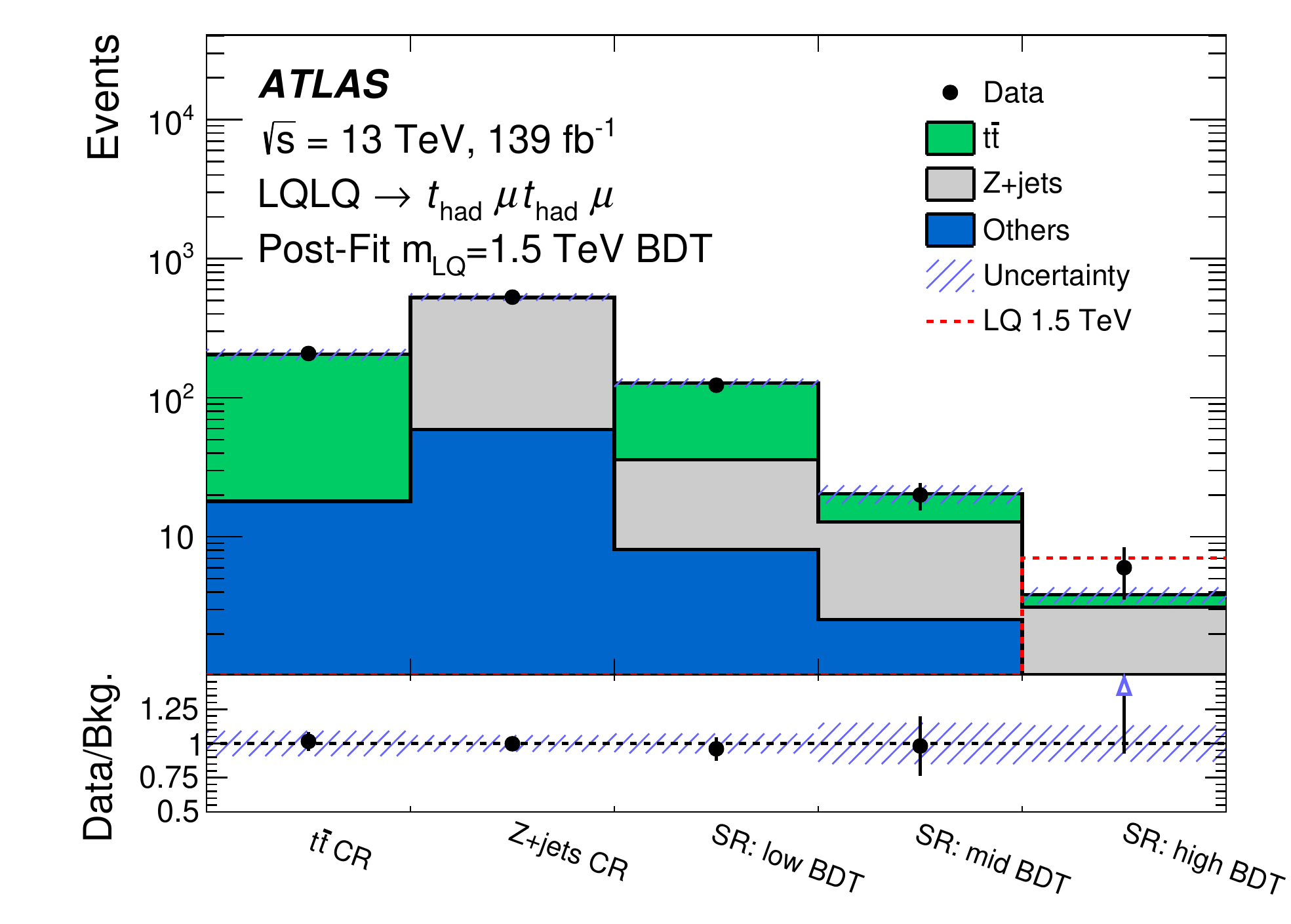}
\caption{Fit results (background-only) for the binned BDT output score distribution in the signal region of the electron (left) and muon (right) channel, and the overall number of events in the \ttbar{} and \zjets{} control regions. The lower panel shows the ratio of data to the fitted background yields. The band represents the systematic uncertainty after the maximum-likelihood fit.}
\label{FigPostFitResult_BONLY}
\end{figure}
 
\subsection{Limits on LQ pair production}\label{Results:limitResults}
 
Upper limits at the 95\% CL on the LQ pair-production cross section, for an assumed value of $\beta=1$, are set as a function of the LQ mass $m_{\mathrm{LQ}}$ and compared with the theoretical prediction (Figure~\ref{fig:FigLimitXS}). The resulting lower limit on $m_{\mathrm{LQ}}$ is determined using the central value of the theoretical NNLO+NNLL cross-section prediction. The observed (expected) lower limits on $m_{\mathrm{LQ}}$ are found to be 1480~(1560)~\GeV\ and 1470~(1540)~\GeV\ for the electron and muon channel respectively. The sensitivity of the analysis is limited by the statistical uncertainty of the data. Including all systematic uncertainties degrades the expected mass limits by only around 10~\GeV, and for a mass of 1.5~\TeV\ the cross-section limits increase by less than 7\% in both the electron and muon channel.
 
Exclusion limits on LQ pair production are also obtained for different values of $m_{\mathrm{LQ}}$ as a function of the branching ratio (\BR) into a charged lepton and a top quark (Figure~\ref{fig:FigLimitGrid}).
The theoretical cross section was scaled by the branching ratio, and then used to obtain the corresponding limit.
The full statistical interpretation is performed for each 0.1 step in \BR, covering the full plane.

\begin{figure}[!hpb]
\begin{center}
\includegraphics[width=0.48\columnwidth]{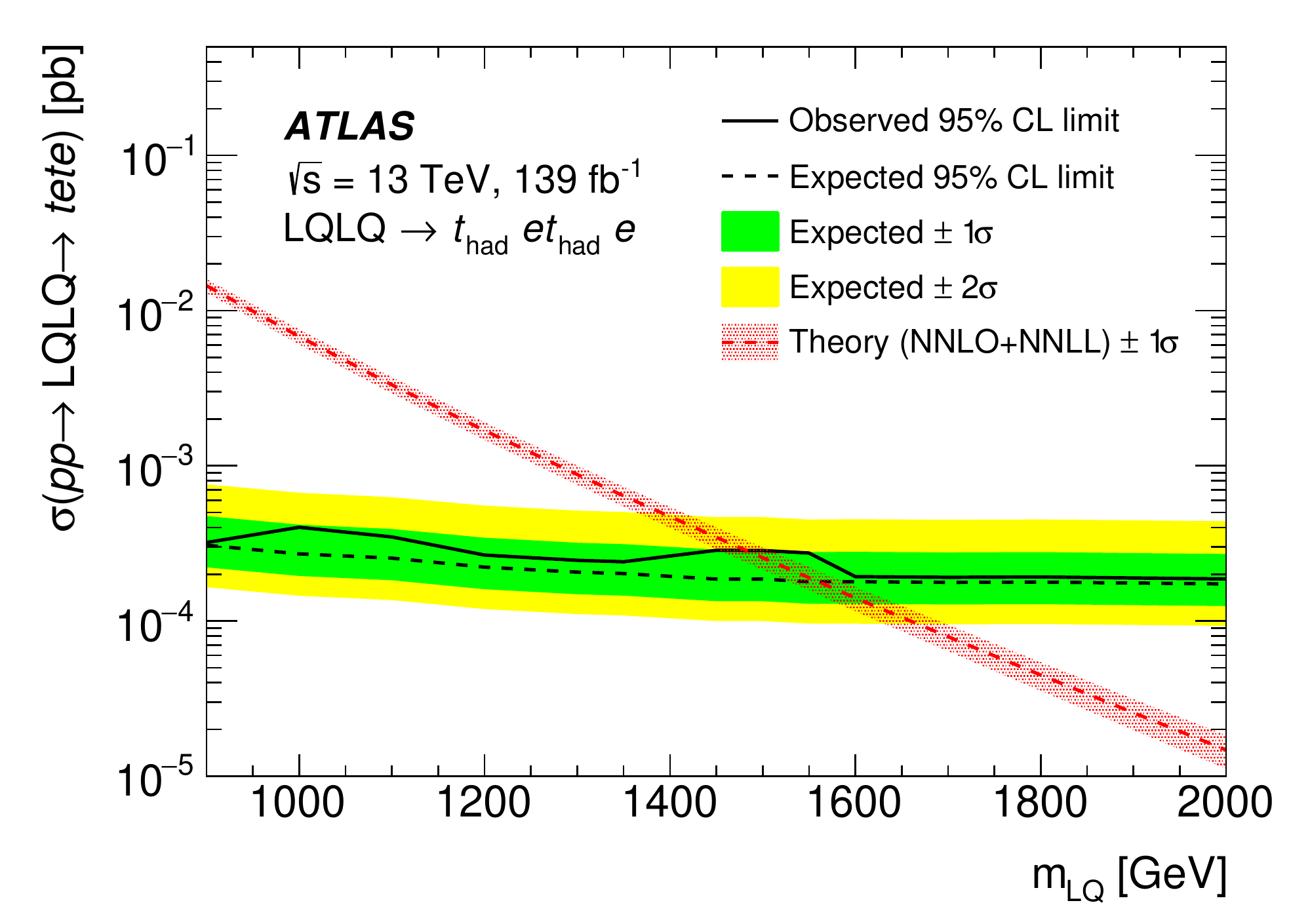}
\includegraphics[width=0.48\columnwidth]{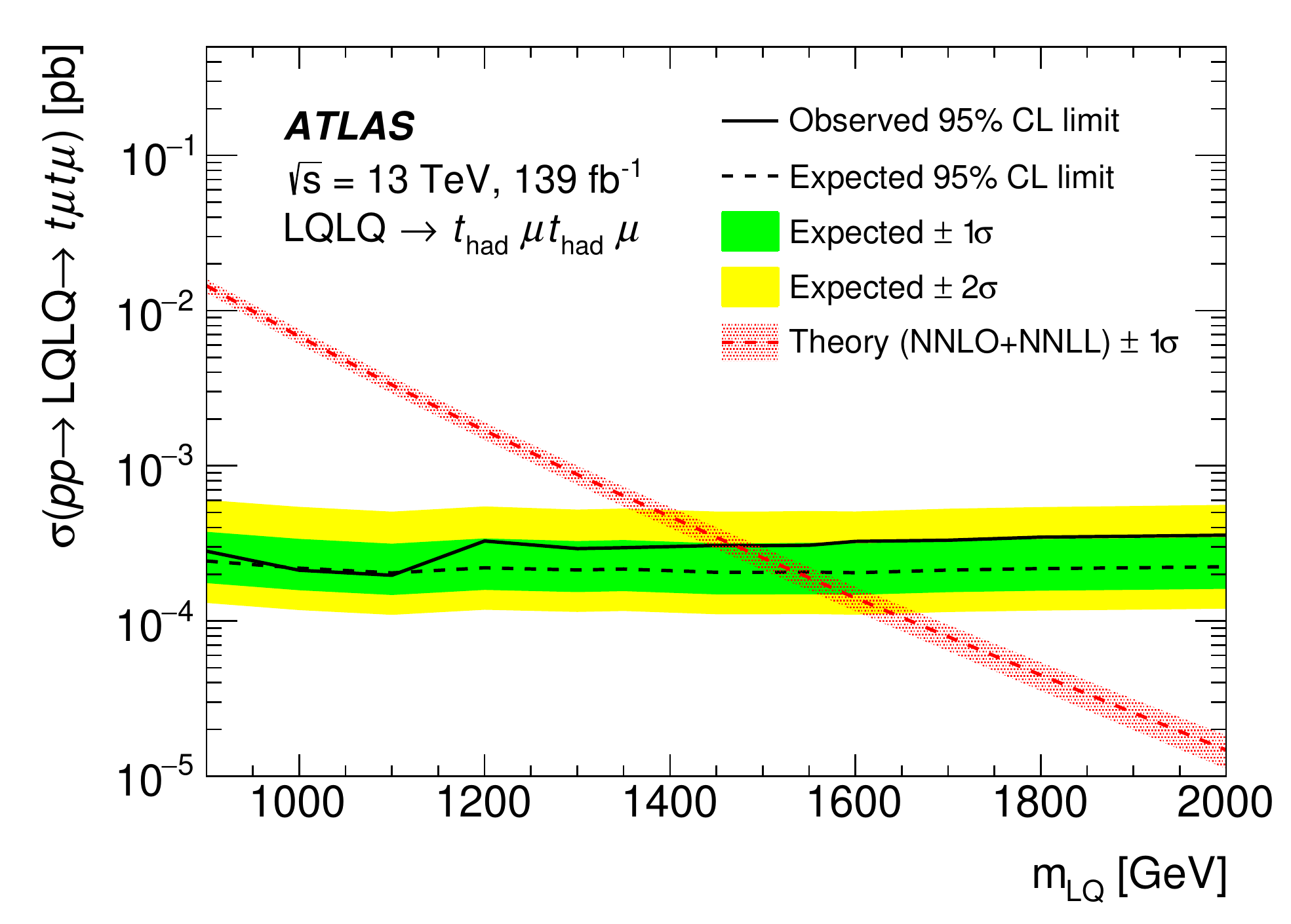}
\caption{Upper limits at 95\% CL on the cross section of LQ pair production as a function of LQ mass, assuming a branching ratio \BR$(\mathrm{LQ} \to t\ell^{\pm}) = 1$, for the electron (left) and muon (right) channel. Observed limits are shown as a black solid line and expected limits as a black dashed line. The green and yellow shaded bands correspond to $\pm$1 and $\pm$2 standard deviations, respectively, around the expected limit. The red curve and band show the nominal theoretical prediction and its $\pm$1 standard deviation uncertainty.}
\label{fig:FigLimitXS}
\end{center}
\end{figure}
 
\begin{figure}[h]
\begin{center}
\includegraphics[width=0.48\columnwidth]{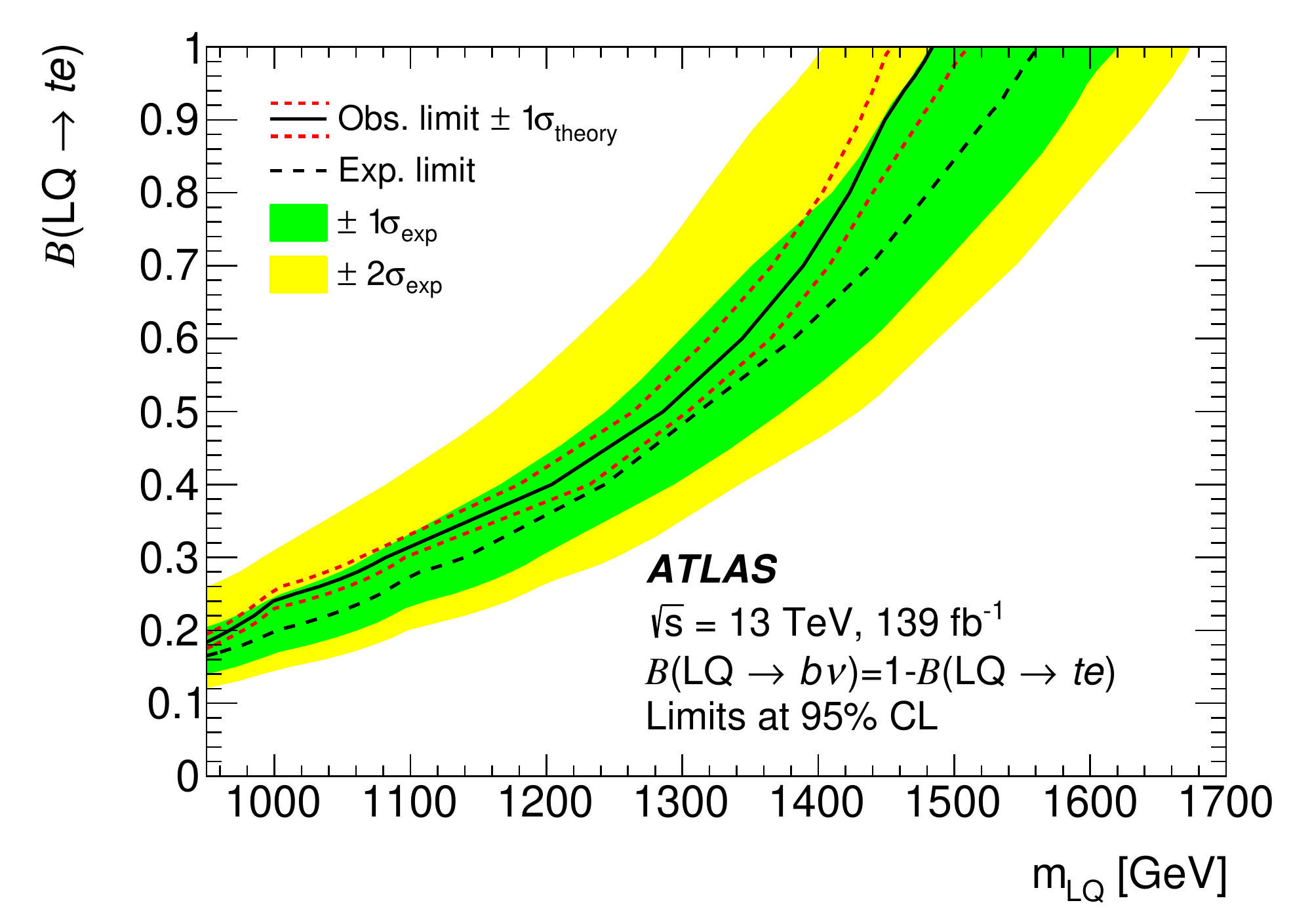}
\includegraphics[width=0.48\columnwidth]{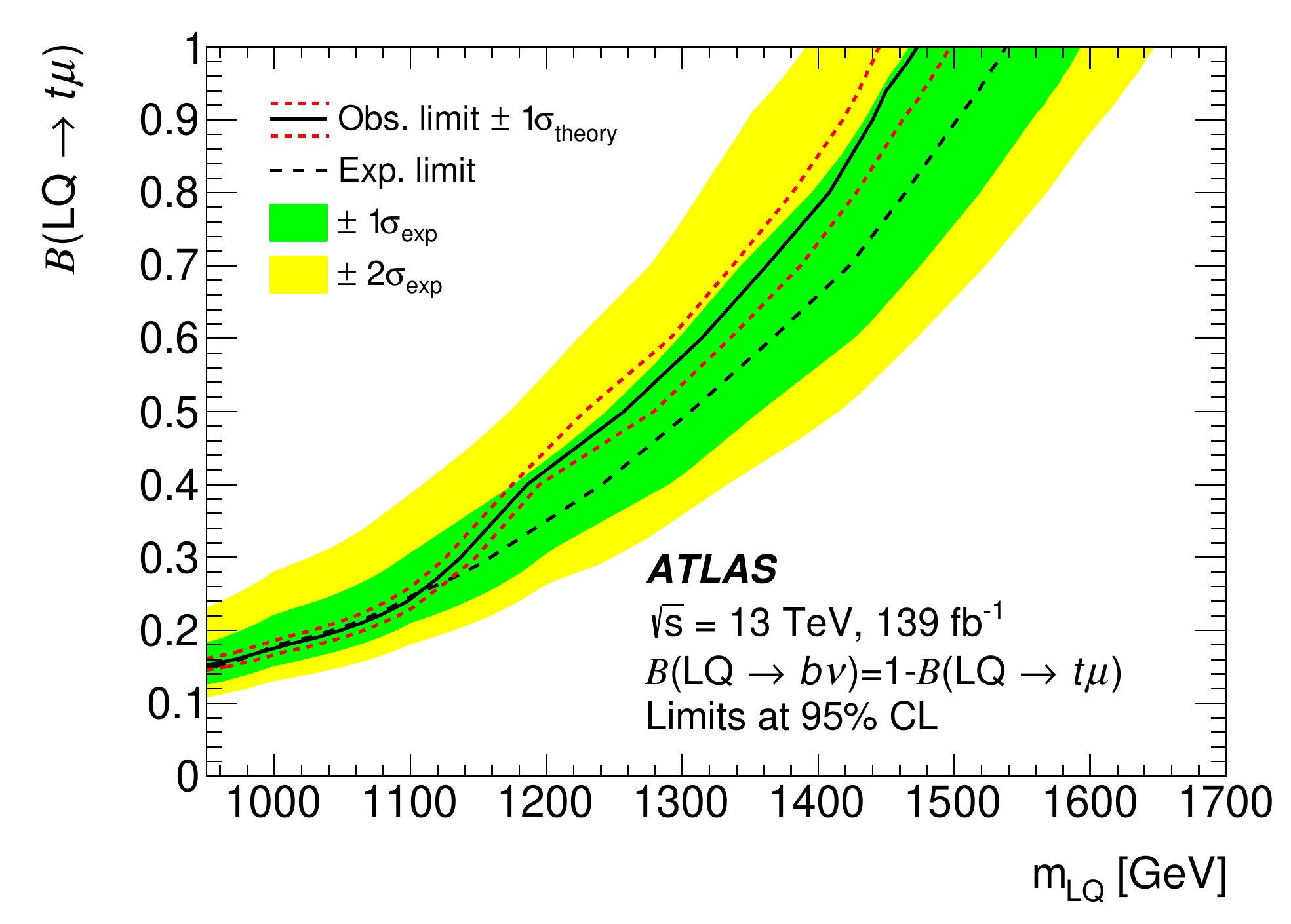}
\caption{Lower exclusion limits on the leptoquark mass for scalar leptoquark pair production as a function of the branching ratio into a top quark and an electron (left) or a muon (right) at 95\% CL. The observed nominal limits are indicated by a black solid curve, with the surrounding red dotted lines obtained by varying the signal cross section by uncertainties from PDFs, renormalization and factorization scales, and the strong coupling constant \alphas. Expected limits are indicated with a black dashed curve, with the yellow and green bands indicating the $\pm$1 standard deviation and $\pm$2 standard deviation excursions due to experimental and modelling uncertainties.}
\label{fig:FigLimitGrid}
\end{center}
\end{figure}
 
\FloatBarrier
\section{Conclusion}
\label{sec:conclusion}
A search for pair production of scalar leptoquarks, each decaying into a top quark and either an electron or a muon has been presented, targeting the high-mass region in which the decay products of each top quark are contained within a single large-radius jet. The analysis is based on tight selection criteria to reduce the SM backgrounds.
The normalizations of the dominant \zjets{} and \ttbar{} backgrounds were determined simultaneously in a profile likelihood fit to the binned output score of a boosted decision tree in the signal region and two dedicated control regions.
The data used in this search correspond to an integrated luminosity of 139~$\mathrm{fb}^{-1}$ of $pp$ collisions with
a centre-of-mass energy $\sqrt{s}=13~\TeV$ recorded by the ATLAS experiment in the whole of Run 2 of the LHC. The observed data distributions are compatible with the expected Standard Model background and no significant excess is observed.
Lower limits on the leptoquark masses are set at 1480~\GeV\ and 1470~\GeV\ for the electron and muon channel, respectively.
\clearpage
\appendix
\part*{Appendix}
\addcontentsline{toc}{part}{Appendix}
\begin{table}[h]
\begin{center}
\caption{
The discriminating variables used in the signal--background discrimination training can be classified into five different groups. The first three groups include kinematic variables that are physics-based rather than detector-based, conditioned on different physics process hypotheses: LQ, dileptonic \ttbar{} and leptonic $Z$ decay hypothesis. These physics-based kinematic variables include the invariant masses and the momenta of intermediate and final-state particles in their parent's rest frame.
In the dileptonic \ttbar{} decay hypothesis, the combinatoric ambiguity in the small-$R$-jet--lepton pairing is resolved using the `min $\Delta M_{\mathrm{top}}$ approach' of the recursive jigsaw reconstruction technique~\cite{RJR}. The reconstructed hemisphere of the decay process associated with the leading (subleading) lepton is labelled with 1 (2). The fourth group of variables is detector-based and defined in the lab frame. These variables are related to the event-level activity of visible objects or missing transverse momentum. The last group of variables is used to identify the three-prong jet structure of hadronic top-quark decays and is used only in the muon channel.
}
\label{tab:discriminating_variables}
\resizebox{0.86\textwidth}{!}{
\centering
\begin{tabular}{l l p{14cm} }
\hline
\multicolumn{3}{l}{Input variables} \\
\toprule
LQ hypothesis & $m_{\text{LQLQ}}$ & Invariant mass of LQ pair system, reconstructed from two leptons and two large-$R$ jets \\
& $m_{\ell\text{1,J1}}^{\text{max}}$ & The higher mass of the two LQ candidates, with the lepton--jet pair labelled as $\ell\text{1 and J1}$. \\
& $m_{\ell\text{2,J2}}^{\text{min}}$ & The lower mass of the two LQ candidates, with the lepton--jet pair labelled as $\ell\text{2 and J2}$. \\
& $m_{\ell\text{2,J1}}$ & Invariant mass of lepton--jet pair $\ell\text{2 and J1}$ \\
& $m_{\ell\text{1,J2}}$ & Invariant mass of lepton--jet pair $\ell\text{1 and J2}$ \\
& $m_{\text{J1}}$ & Invariant mass of large-$R$ jet J1 \\
& $m_{\text{J2}}$ & Invariant mass of large-$R$ jet J2 \\
& $E_{\ell1}^{\text{LQ}}$ & Energy of lepton $\ell\text{1}$ in its LQ parent's rest frame \\
& $E_{\ell2}^{\text{LQ}}$ & Energy of lepton $\ell\text{2}$ in its LQ parent's rest frame \\
& $E_{t1}^{\text{LQ}}$ & Energy of large-$R$ jet J1 in its LQ parent's rest frame \\
& $E_{t2}^{\text{LQ}}$ & Energy of large-$R$ jet J2 in its LQ parent's rest frame \\
Dilepton \ttbar{} hypothesis & $m_{tt}$ & Invariant mass of \ttbar{} system, reconstructed from two leptons, two resolved jets and $E_\mathrm{T}^\mathrm{miss}$ \\
& $m_{t1'}$ & Invariant mass of top quark $t1'$, reconstructed from $W$ boson $W1'$ and $b$-quark $b1'$\\
& $m_{t2'}$ & Invariant mass of top quark $t2'$, reconstructed from $W$ boson $W2'$ and $b$-quark $b2'$\\
& $m_{t1'\text{, swapped}}$ & Invariant mass of top quark $t1'$, with its $b$-quark child $b1'$ swapped with that of top quark $t2'$ \\
& $m_{t2'\text{, swapped}}$ & Invariant mass of top quark $t2'$, with its $b$-quark child $b2'$ swapped with that of top quark $t1'$ \\
& $m_{W1'}$ & Invariant mass of $W$ boson $W1'$, reconstructed from the leading lepton $\ell1'$ and $E_\mathrm{T}^\mathrm{miss}$ \\
& $m_{W2'}$ & Invariant mass of $W$ boson $W2'$, reconstructed from the subleading lepton $\ell2'$ and $E_\mathrm{T}^\mathrm{miss}$ \\
& $E_{b1'}^{t}$ & Energy of small-$R$ jet j1 as $b$-quark candidate $b1'$ in its top quark parent ($t1'$) rest frame \\
& $E_{b2'}^{t}$ & Energy of small-$R$ jet j2 as $b$-quark candidate $b2'$ in its top quark parent ($t2'$) rest frame \\
& $E_{\ell1'}^{W}$ & Energy of the leading lepton $\ell1'$ in its $W$ boson parent ($W1'$) rest frame \\
& $E_{\ell2'}^{W}$ & Energy of the subleading lepton $\ell2'$ in its $W$ boson parent ($W2'$) rest frame \\
$Z\rightarrow\ell\ell$ hypothesis & $m_{\ell\ell}$ & Invariant mass of the dilepton system\\
& $p_{\text{T,}\ell\ell}^{\text{lab}}$ & Transverse momentum of the dilepton system in the lab frame\\
Detector-based & $L_{\text{T}}$ & Scalar $p_{\text{T}}$ sum of the two leptons \\
& $H_{\text{T}}$ & Scalar $p_{\text{T}}$ sum of the two leading large-$R$ jets\\
& $S_{\text{T}}$ & Scalar $p_{\text{T}}$ sum of the two leptons and the two leading large-$R$ jets\\
& \MET & Missing transverse momentum \\
& \MET{} sig.& Missing transverse momentum significance, defined as $\MET/\sqrt{H_{\text{T}}}$\\
Jet substructure & $\text{sd}_{23}$ & $k_t$ splitting scale for the 2nd and 3rd subjet, defined as $\text{sd}_{23}=\text{min}(p_{\text{T},2},p_{\text{T},3})\times \Delta R_{23}$ \\
& $\tau_{32}^{\text{WTA}}$ & The ratio of $\tau_{3}$ to $\tau_{2}$, where N-subjettiness variable $\tau_{N}$ is defined as $\tau_{N}=\frac{1}{d_{0}}\sum_{i\in \text{jet constituents}} p_{\text{T},i}\times \text{min}(\delta R_{1i}, ..., \delta R_{Ni})$ with $d_{0}=\sum_{i\in \text{jet constituents}} p_{\text{T},i}\times R$, where $R$ is the radius parameter of the jet, and $\delta R_{ji}$ is the distance between the subjet j and the constituent $i$. WTA denotes the winner-take-all (WTA) recombination scheme~\cite{wta} used in subjet reconstruction. \\
& $Q_{\text{w}}$ & The minimum invariant mass of the two subjets in the second-to-last reclustering step of the $k_t$ algorithm, applied to a large-$R$ jet \\
MVA parameterization & $m_{\text{LQ, hypo}}$ & Set to the test mass point at which the model is utilized. In the training phase, this parameter is set to the corresponding LQ mass for the signal samples, and a uniformly distributed random value from the training set of LQ mass points for the background samples.\\
\hline
\end{tabular}
}
\end{center}
\end{table}
 
\FloatBarrier
\clearpage
 
\section*{Acknowledgements}

We thank CERN for the very successful operation of the LHC, as well as the
support staff from our institutions without whom ATLAS could not be
operated efficiently.
 
We acknowledge the support of ANPCyT, Argentina; YerPhI, Armenia; ARC, Australia; BMWFW and FWF, Austria; ANAS, Azerbaijan; SSTC, Belarus; CNPq and FAPESP, Brazil; NSERC, NRC and CFI, Canada; CERN; ANID, Chile; CAS, MOST and NSFC, China; COLCIENCIAS, Colombia; MSMT CR, MPO CR and VSC CR, Czech Republic; DNRF and DNSRC, Denmark; IN2P3-CNRS and CEA-DRF/IRFU, France; SRNSFG, Georgia; BMBF, HGF and MPG, Germany; GSRT, Greece; RGC and Hong Kong SAR, China; ISF and Benoziyo Center, Israel; INFN, Italy; MEXT and JSPS, Japan; CNRST, Morocco; NWO, Netherlands; RCN, Norway; MNiSW and NCN, Poland; FCT, Portugal; MNE/IFA, Romania; JINR; MES of Russia and NRC KI, Russian Federation; MESTD, Serbia; MSSR, Slovakia; ARRS and MIZ\v{S}, Slovenia; DST/NRF, South Africa; MICINN, Spain; SRC and Wallenberg Foundation, Sweden; SERI, SNSF and Cantons of Bern and Geneva, Switzerland; MOST, Taiwan; TAEK, Turkey; STFC, United Kingdom; DOE and NSF, United States of America. In addition, individual groups and members have received support from BCKDF, CANARIE, Compute Canada, CRC and IVADO, Canada; Beijing Municipal Science \& Technology Commission, China; COST, ERC, ERDF, Horizon 2020 and Marie Sk{\l}odowska-Curie Actions, European Union; Investissements d'Avenir Labex, Investissements d'Avenir Idex and ANR, France; DFG and AvH Foundation, Germany; Herakleitos, Thales and Aristeia programmes co-financed by EU-ESF and the Greek NSRF, Greece; BSF-NSF and GIF, Israel; La Caixa Banking Foundation, CERCA Programme Generalitat de Catalunya and PROMETEO and GenT Programmes Generalitat Valenciana, Spain; G\"{o}ran Gustafssons Stiftelse, Sweden; The Royal Society and Leverhulme Trust, United Kingdom.
 
The crucial computing support from all WLCG partners is acknowledged gratefully, in particular from CERN, the ATLAS Tier-1 facilities at TRIUMF (Canada), NDGF (Denmark, Norway, Sweden), CC-IN2P3 (France), KIT/GridKA (Germany), INFN-CNAF (Italy), NL-T1 (Netherlands), PIC (Spain), ASGC (Taiwan), RAL (UK) and BNL (USA), the Tier-2 facilities worldwide and large non-WLCG resource providers. Major contributors of computing resources are listed in Ref.~\cite{ATL-SOFT-PUB-2020-001}.
 

\printbibliography

\clearpage \input{atlas_authlist}

\end{document}

%% file: atlas_authlist.tex
 
\begin{flushleft}
\hypersetup{urlcolor=black}
{\Large The ATLAS Collaboration}

\bigskip

\AtlasOrcid[0000-0002-6665-4934]{G.~Aad}$^\textrm{\scriptsize 102}$,    
\AtlasOrcid[0000-0002-5888-2734]{B.~Abbott}$^\textrm{\scriptsize 128}$,    
\AtlasOrcid[0000-0002-7248-3203]{D.C.~Abbott}$^\textrm{\scriptsize 103}$,    
\AtlasOrcid[0000-0002-2788-3822]{A.~Abed~Abud}$^\textrm{\scriptsize 36}$,    
\AtlasOrcid[0000-0002-1002-1652]{K.~Abeling}$^\textrm{\scriptsize 53}$,    
\AtlasOrcid[0000-0002-2987-4006]{D.K.~Abhayasinghe}$^\textrm{\scriptsize 94}$,    
\AtlasOrcid[0000-0002-8496-9294]{S.H.~Abidi}$^\textrm{\scriptsize 167}$,    
\AtlasOrcid[0000-0002-8279-9324]{O.S.~AbouZeid}$^\textrm{\scriptsize 40}$,    
\AtlasOrcid{N.L.~Abraham}$^\textrm{\scriptsize 156}$,    
\AtlasOrcid[0000-0001-5329-6640]{H.~Abramowicz}$^\textrm{\scriptsize 161}$
% ...
\bigskip
\\

% ...
$^{36}$CERN, Geneva; Switzerland.\\
% ...
$^{40}$Niels Bohr Institute, University of Copenhagen, Copenhagen; Denmark.\\
% ...
$^{53}$II. Physikalisches Institut, Georg-August-Universit\"{a}t G\"ottingen, G\"ottingen; Germany.\\
% ...
$^{94}$Department of Physics, Royal Holloway University of London, Egham; United Kingdom.\\
% ...
$^{102}$CPPM, Aix-Marseille Universit\'e, CNRS/IN2P3, Marseille; France.\\
$^{103}$Department of Physics, University of Massachusetts, Amherst MA; United States of America.\\
% ...
$^{128}$Homer L. Dodge Department of Physics and Astronomy, University of Oklahoma, Norman OK; United States of America.\\
% ...
$^{156}$Department of Physics and Astronomy, University of Sussex, Brighton; United Kingdom.\\
% ...
$^{161}$Raymond and Beverly Sackler School of Physics and Astronomy, Tel Aviv University, Tel Aviv; Israel.\\
% ...
$^{167}$Department of Physics, University of Toronto, Toronto ON; Canada.\\
% ...

\end{flushleft}
